\documentclass[titlepage,12pt]{article}
\usepackage{hyperref}
\usepackage{amssymb,amsmath,color,graphics,amscd,epsf,indentfirst,amsfonts}
\usepackage[square,comma,sort&compress]{natbib}
\usepackage{hypernat}
\usepackage{epsfig}
\usepackage{nicefrac}
\usepackage{scalefnt}
\usepackage[titles]{tocloft}
\usepackage{sectsty}
\allsectionsfont{\bf \scalefont{.7} \selectfont}
\subsectionfont{\bf \scalefont{.85} \it \selectfont}
\subsubsectionfont{\bf \scalefont{1} \it \selectfont}
\usepackage[T1]{fontenc}
\usepackage{lmodern}
\usepackage{bibspacing}

\def\blfootnote{\xdef\@thefnmark{}\@footnotetext}

\long\def\symbolfootnote[#1]#2{\begingroup%
\def\thefootnote{\fnsymbol{footnote}}\footnote[#1]{#2}\endgroup}

\makeatletter
\renewcommand{\@dotsep}{4.5}
\makeatother

\def\be{\begin{equation}}
\def\ee{\end{equation}}

\makeatletter
\def\@seccntformat#1{\csname the#1\endcsname.\quad}
\makeatother

\def\clock{{\count0=\time
           \divide\count0 60
           \ifnum\count0<10 0\fi\the\count0
           \multiply\count0 -60 \advance\count0 \time
           :\ifnum\count0<10 0\fi \the\count0
         }}
\newcommand{\timestamp}{{\small\vbox{\hbox{\tt\jobname.tex}
\hbox{\the\day/\the\month/\the\year, \clock}}}}

\def\LL{{\cal L}}
\def\MM{{\cal M}}
\def\NN{{\cal N}}

\def\SS{{\cal S}}

\def\d{{\partial}}

\newcommand{\beq}{\begin{equation}}
\newcommand{\eeq}{\end{equation}}
\newcommand{\ba}{\begin{array}}
\newcommand{\ea}{\end{array}}
\newcommand{\bea}{\begin{eqnarray}}
\newcommand{\eea}{\end{eqnarray}}

\newcommand{\Z}{\mathbb{Z}}

\newcommand{\tr}{\mathop{{\rm Tr}}}
\DeclareMathOperator{\sgn}{sgn}

\numberwithin{equation}{section}

\begin{document}

\begin{titlepage}
\begin{flushright}
CPHT-RR011.0209\\
\vskip -1cm
\end{flushright}
\vskip 2.8cm
\begin{center}
\font\titlerm=cmr10 scaled\magstep4
    \font\titlei=cmmi10 scaled\magstep4
    \font\titleis=cmmi7 scaled\magstep4
    \centerline{\titlerm
      R-charges, Chiral Rings and RG Flows in}
      \vspace{0.4cm}
    \centerline{\titlerm
       Supersymmetric Chern-Simons-Matter Theories}   
\vskip 1.5cm
{Vasilis Niarchos}\\
\vskip 0.7cm
\medskip
{\it Centre de Physique Th\'eorique, \'Ecole Polytechnique,
91128 Palaiseau, France}\\
{\it Unit\'e mixte de Recherche 7644, CNRS}\\
\smallskip
{niarchos@cpht.polytechnique.fr}

\end{center}
\vskip .4in
\centerline{\bf Abstract}

\baselineskip 20pt
%

\vskip .5cm \noindent
We discuss the non-perturbative behavior of the $U(1)_R$ symmetry in 
$\NN=2$ superconformal Chern-Simons theories coupled to matter in the 
(anti)fundamental and adjoint representations of the gauge group, which we
take to be $U(N)$. Inequalities constraining this behavior are obtained as 
consequences of spontaneous breaking of supersymmetry and Seiberg duality. 
This information reveals a web of RG flows connecting different interacting
superconformal field theories in three dimensions. We observe that a subclass 
of these theories admits an ADE classification. In addition, we postulate new 
examples of Seiberg duality in $\NN=2$ and $\NN=3$ Chern-Simons-matter 
theories and point out interesting parallels with familiar non-perturbative 
properties in $\NN=1$ (adjoint) SQCD theories in four dimensions where 
the exact $U(1)_R$ symmetry can be determined using $a$-maximization.

\vfill
\noindent
March 2009
\end{titlepage}\vfill\eject

\setcounter{equation}{0}

\pagestyle{empty}
\small
\vspace*{-0.7cm}
\tableofcontents
\normalsize
\newpage
\pagestyle{plain}
\setcounter{page}{1}

\section{Introduction}
\label{sec:intro}

Three-dimensional gauge theories exhibit a range of interesting, non-perturbative 
phenomena (see, for example, \cite{Intriligator:1996ex,Hanany:1996ie,deBoer:1997kr,
Aharony:1997bx,Karch:1997ux,Aharony:1997gp} and references/citations thereof). 
Some of the complicating features of these theories stem from the fact that the gauge 
interaction is a classically relevant operator in three dimensions. As we flow towards 
the infrared (IR), the gauge coupling grows indefinitely and the theory becomes 
strongly coupled. A way to ameliorate this strong coupling problem is to add to the 
Lagrangian the Chern-Simons (CS) interaction
\beq
\label{introaa}
\SS_{\rm CS}=\frac{k}{4\pi}\int \tr\left (A\wedge dA+\frac{2}{3} A\wedge A\wedge A\right)
\eeq
where $A$ is the gauge field one-form and the constant $k$ is the CS level. 
This interaction, which is specific to three dimensions, makes the gauge field massive 
with a mass of order
\beq
\label{introab}
m_{\rm CS}\sim g_{\rm YM}^2 k
\eeq
where $g_{\rm YM}$ is the gauge coupling. At energies below the scale set by $m_{\rm CS}$
the IR behavior of the theory is controlled by the CS interaction and the flow towards strong 
coupling is effectively cutoff. 

Chern-Simons theories coupled to matter are non-trivial quantum field theories. Without 
matter the Chern-Simons Lagrangian \eqref{introaa} defines a topological quantum field 
theory with a fascinating connection to two-dimensional Wess-Zumino-Witten (WZW) models 
\cite{Witten:1988hf}.

In this paper we will discuss Chern-Simons-Matter (CSM) theories with at least four 
supercharges, that is $\NN=2$ supersymmetry in three dimensions. These theories
are characterized by a gauge group $G$, the CS level $k$ and the representations
$R_i$ of the matter fields. The gauge field is part of the $\NN=2$ vector multiplet $V$ 
and matter is organized in $\NN=2$ chiral multiplets $\Phi_i$. The components of 
$\NN=2$ multiplets can be deduced easily by dimensional reduction from $\NN=1$ 
multiplets in four dimensions. The vector multiplet contains the three-dimensional 
gauge field $A_\mu$, a scalar $\sigma$, an auxiliary scalar $D$ and a two-component 
Dirac spinor $\chi$. A chiral multiplet $\Phi_i$ contains a complex scalar $\varphi_i$ 
and a Dirac spinor $\psi_i$.

The $\NN=2$ supersymmetric CS action has a known expression in superspace
language \cite{Zupnik:1988en,Ivanov:1991fn,Avdeev:1991za,Avdeev:1992jt}. 
This expression becomes simpler in Wess-Zumino (WZ) gauge
\beq
\label{introac}
\SS_{\rm CS}^{\NN=2}=\frac{k}{4\pi}\int \tr
\left(A\wedge dA+\frac{2}{3} A\wedge A\wedge A-\bar \chi \chi+2D\sigma
\right)
~.
\eeq
The matter multiplets have the standard kinetic terms
\beq
\label{introad}
\SS_{\rm matter, kin}=\int d^3x\, d^4\theta\, \sum_i \bar \Phi_i e^{V}\Phi_i
~.
\eeq
Further coupling via superpotential interactions is possible.

Integrating out the auxiliary fields and massive fermions of the $\NN=2$ vector 
multiplet gives an interacting theory of scalars and fermions \cite{Gaiotto:2007qi}. 
Besides their intrinsic interest, such theories have important applications in 
M-theory and the AdS/CFT correspondence. Conformal CSM theories with 
enhanced supersymmetry ($\NN=4,5,6,8$) have been argued \cite{Aharony:2008ug,
Hosomichi:2008jd,Hosomichi:2008jb,Schnabl:2008wj,Aharony:2008gk} 
to provide a gauge theory description of the infrared dynamics of M2-branes 
in different setups and to be related by holography to string/M-theory on AdS$_4$ 
backgrounds. In this paper, our main objective are the field theory properties of
$\NN=2$ CSM theories, although some comments will be made on aspects 
related to M-brane dynamics and the AdS$_4$/CFT$_3$ correspondence.

Assuming the absence of superpotential interactions, all couplings in the
action 
\beq
\label{introad}
\SS_{\rm CSM}^{\NN=2}=\SS_{\rm CS}^{\NN=2}+\SS_{\rm matter, kin}
\eeq 
are controlled by the inverse of the CS level $\frac{1}{k}$. The theory is 
weakly coupled at large $k$. For non-abelian gauge groups $G$ the level $k$ 
is an integer, and does not receive quantum corrections, except for a possible 
one-loop shift \cite{Avdeev:1992jt,Kapustin:1994mt,Witten:1999ds}. In fact,
one can argue that these theories are both classically and quantum mechanically 
superconformal \cite{Gaiotto:2007qi}. In these superconformal theories the 
$U(1)_R$ symmetry is non-trivial and depends on $k$. In other words, the 
scaling dimensions of chiral operators receive $k$-dependent anomalous 
dimensions.

The addition of relevant superpotential interactions to the action \eqref{introad}
breaks the conformal invariance and generates a renormalization group (RG) 
flow towards new interacting fixed points. Identifying these flows and determining 
analytically their properties is, in general, a non-perturbative question that remains 
largely open.

In order to make progress in this problem it is desirable to determine with 
exact analytical methods the $U(1)_R$ symmetry in any of the above fixed 
points. Knowing this symmetry would allow us to determine the exact scaling
dimension of chiral operators. Relevant chiral operators can be added to 
the Lagrangian as superpotential deformations to generate new RG flows 
and IR fixed points. 

A similar question can be posed in four-dimensional gauge theories with 
$\NN=1$ supersymmetry. In this context, the exact $U(1)_R$ symmetry can 
be determined with a combination of $a$-maximization and 't Hooft anomaly 
matching \cite{Intriligator:2003jj} .

There are several tools that allow us to compute non-perturbative 
quantities in four-dimensional $\NN=1$ gauge theories. Many of them are 
closely related to the presence of anomalies. The NSVZ exact $\beta$-function
formula \cite{Novikov:1983uc}, $a$-maximization \cite{Intriligator:2003jj} and 
't Hooft anomaly matching are well known examples. In some cases, additional
information can be obtained with the use of Seiberg duality \cite{Seiberg:1994pq},
which is a powerful strong/weak coupling duality. Since there are no anomalies of 
continuous symmetries in three dimensions a corresponding understanding of the 
properties of three-dimensional quantum field theories is currently lacking.  

In an effort to obtain a more precise understanding of the properties of 
$U(1)_R$ symmetries and RG flows in $\NN=2$ CSM theories, we will 
examine in this paper what happens in $U(N_c)$ $\NN=2$ CS theories 
coupled to a set of matter fields that contains: $N_f$ chiral superfields 
$Q^i$ $(i=1,\cdots, N_f)$ in the fundamental representation, $N_f$ chiral 
superfields $\widetilde Q_i$ in the antifundamental and zero, one or two 
chiral superfields in the adjoint representation of the gauge group. Non-trivial 
RG flows can be generated in these theories with superpotential deformations 
that involve gauge invariant chiral operators made out of the above fields.

In the presence of superpotential interactions these theories can be viewed as 
three-dimensional versions of $\NN=(2,2)$ Landau-Ginzburg (LG) models in two 
dimensions. They can also be viewed as three-dimensional versions of $\NN=1$ 
SQCD theories in four dimensions with $N_f$ flavor chiral multiplets and zero, one 
or two adjoints.\footnote{In four dimensions two is the maximum number of adjoints 
if we require asymptotic freedom. We do not have a corresponding restriction 
in three dimensions, but since we want to compare the properties of three and 
four-dimensional gauge theories we will also restrict our discussion to CSM 
theories with up to two adjoint chiral superfields.} Because of the many similarities 
with the four-dimensional (adjoint) SQCD theories we will sometimes call the 
corresponding CSM theories CS-SQCD, 1-adjoint CS-SQCD and 2-adjoint 
CS-SQCD.

We will find that the similarities between CS-SQCD and SQCD theories in
three and four dimensions respectively are not limited to the matter content but 
extend to non-perturbative aspects of their dynamics. Some of the 
common features can be traced back to the similarities between the chiral 
rings. Other features, however, like the stability of the supersymmetric vacuum 
and Seiberg duality, are highly non-trivial and appear to arise in three and four 
dimensions through different mechanisms. These similarities are impressive 
and reveal how rich the dynamics of $\NN=2$ CSM theories is.

Spontaneous supersymmetry breaking and Seiberg duality are properties that 
arise in some of the theories we will examine. They can be inferred most easily 
from the rules of brane dynamics in configurations of D-branes and NS5-branes in 
type IIB string theory \cite{Bergman:1999na,Ohta:1999iv,Giveon:2008zn,Niarchos:2008jb}. 
In some cases, where a string theory construction is unknown, we will argue for 
bounds of spontaneous breaking of supersymmetry using the chiral ring structure 
directly in field theory. These properties have important consequences for the
exact $U(1)_R$ symmetry in these theories and as such they will provide a useful
semi-quantitative guide to the behavior of R-charges as we move from the weak 
to the strong coupling regime. This indirect reasoning is what will allow us to
make some progress despite our lack of an exact analytic tool in three
dimensions like $a$-maximization.

With this information about R-symmetries we will be able to identify a web of
RG flows connecting different conformal field theories (CFTs) in three dimensions. 
In the IR of these flows interacting fixed points of the $\beta$-function arise from a 
balancing of two counteracting sources: the gauge interactions that work to decrease
the R-charges and the superpotential interactions that work to increase them.
This can be verified explicitly in some cases with a two-loop computation in 
perturbation theory. Furthermore, we find that a subset of the CFTs arising in 
this way admits an ADE classification. A similar classification was also observed 
in four-dimensional two-adjoint SQCD theories \cite{Intriligator:2003mi}. 

The organization of this paper is as follows. Section \ref{sec:CSSQCD} reviews 
the key results obtained in Ref.\ \cite{Giveon:2008zn} about CS-SQCD theories 
and supplements them with new observations on the phase structure and 
the $U(1)_R$ symmetry. Section \ref{sec:SQCDsuperpot} considers RG flows that
arise in CS-SQCD theories with superpotential deformations that involve quadratic and
cubic meson interactions. Section \ref{sec:1adj} is devoted to the 1-adjoint CS-SQCD 
theories reviewing and extending the results obtained in Ref.\ \cite{Niarchos:2008jb}. 
Special emphasis is given to the R-charge of the adjoint chiral superfield as a function 
of the parameters of the theory. Sections \ref{sec:hato} and \ref{sec:hatom} discuss RG 
flows that arise in 2-adjoint CS-SQCD theories with superpotential deformations that 
involve single trace, and in some cases also mesonic, chiral operators. In the process, 
we encounter a new set of Chern-Simons theories that exhibit Seiberg duality and the 
emergence of a partial ADE classification of fixed points. We conclude with an overall 
discussion and a list of interesting open problems in section \ref{sec:discussion}.
Some useful technical details are relegated to two appendices at the end of the paper.

\section{CS-SQCD}
\label{sec:CSSQCD}

\vspace{-.3cm}
\subsection{Definition and phase structure}
\label{subsec:defSQCD}

The first set of theories in our agenda are Chern-Simons versions of the 
four-dimensional $\NN=1$ SQCD theories. The Lagrangian that defines these
theories consists of the $\NN=2$ CS interaction at level $k$ and the $\NN=2$
kinetic terms for $N_f$ pairs of $\NN=2$ chiral multiplets $Q^i$, $\widetilde Q_i$
($i=1,2,\cdots,N_f$) in the fundamental (resp.\ antifundamental) representation 
of the gauge group. Explicit expressions for this Lagrangian can be found, for 
example, in Ref.\ \cite{Gaiotto:2007qi}. We will consider the unitary gauge group 
$G=U(N_c)$. In contrast to what happens in four dimensions, here the $U(1) 
\subset U(N_c)$ is interacting and will have some important implications that will 
be discussed below.

We will restrict the level $k$ to be positive. Negative values of $k$ can be obtained
by a parity transformation $x^\mu \to -x^\mu$ that effectively sends $k\to -k$. One 
should think of $1/k$ as the `gauge coupling' of the CS theory. In the large $N_c$ 
limit, the ratio $\lambda=\frac{N_c}{k}$ plays the r\^ole of the 't Hooft coupling. For 
simplicity, in what follows, we will consider our theories in the large-$N$ limit where 
$k, N_c, N_f\gg 1$ with the ratios $\lambda$ and $x=\frac{N_c}{N_f}$ kept finite. This is 
not necessary for most of the statements that we make below. When finite-$N$ effects
do not modify the qualitative picture it will be more convenient to think in terms of 
continuous, instead of discrete, parameters.

The so-defined CS-SQCD theories are superconformal both classically and 
quantum mechanically \cite{Gaiotto:2007qi}. The global symmetry of the theory 
is $SU(N_f)\times SU(N_f) \times U(1)_a\times U(1)_R$.

As we increase the number of colors $N_c$, say for fixed $k$ and $N_f$, we encounter 
a critical point where supersymmetry gets spontaneously broken. At the critical point
\beq
\label{defSQCDaa}
N_c=N_f+k
~.
\eeq

\begin{figure}[t!]
\vspace{-1cm}
\centering
\includegraphics[width=15cm, height=10.5cm]{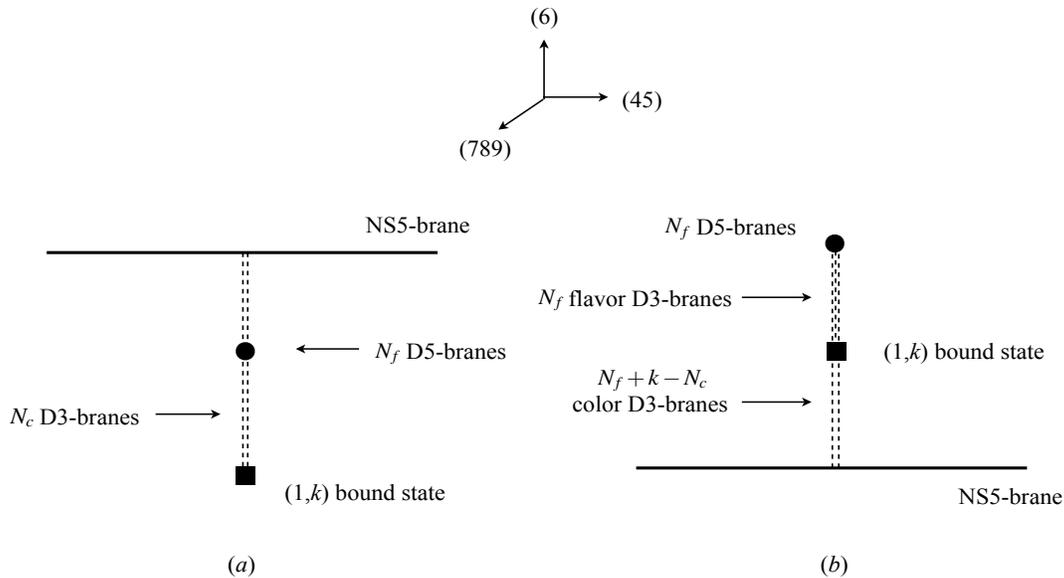}
\vspace{-1.1cm}\bf
\caption{\it Configuration (a) engineers the electric version of the
$\NN=2$ CS-SQCD theory. Configuration (b) engineers the magnetic
version.}
\label{CSSQCD}
\end{figure}

The simplest way to obtain this spontaneous breaking of supersymmetry
is by embedding the theory in a configuration of D-branes and NS5-branes
in type IIB string theory. The relevant setup was analyzed in Ref.\ \cite{Giveon:2008zn}
and we briefly review it here for completeness. It consists of (see also the 
configuration $(a)$ in Fig.\ \ref{CSSQCD})
\beq
\label{defSQCDab}
\begin{array}{r c l}
1~ NS5 ~&:&~~ 0~1~2~3~4~5
\\[1mm]
1~ (1,k)~&:&~~ 0~1~2~\left[ {3 \atop 7} \right]_\theta~ 8~9
\\[1mm]
N_c~ D3~&:&~~ 0~1~2~|6|
\\[1mm]
N_f~ D5~&:&~~ 0~1~2~7~8~9
\end{array}
\eeq
$(1,k)$ denotes the bound state of one NS5-brane and $k$ D5-branes. 
$\left[ {3 \atop 7} \right]_\theta$ denotes an orientation in the (37) plane
at an angle $\theta$ with respect to the axis 3. In the above configuration
$\theta$ is an angle fixed by $k$ by the relation \cite{Aharony:1997ju}
\beq
\label{defSQCDac}
\tan \theta=g_s k
~.
\eeq
$|6|$ denotes that the D3-branes have a finite length along the $x^6$ 
direction. The CS-SQCD theory arises in this setup as the effective low-energy 
description of the dynamics of the theory that resides in the three-dimensional 
intersection of this configuration \cite{Kitao:1998mf}.

In this context, the condition for the existence of a supersymmetric vacuum is a 
consequence of the $s$-rule of brane dynamics \cite{Bergman:1999na,Ohta:1999iv}. 
In Fig.\ \ref{CSSQCD}(a) the $s$-rule constrains the number of D3-branes stretching 
directly between the NS5-brane and the $(1,k)$ bound state. This number $(N_c-N_f)$ 
must be smaller or equal to $k$ in order to preserve supersymmetry, hence the condition 
\beq
\label{defSQCDad}
N_c\leq N_f+k
~.
\eeq
In terms of $\lambda$ and $x$ this condition reads
\beq
\label{defSQCDae}
\lambda \leq \frac{x}{x-1}
~.
\eeq

\begin{figure}[t!]
\centering
\includegraphics[width=13cm, height=8.5cm]{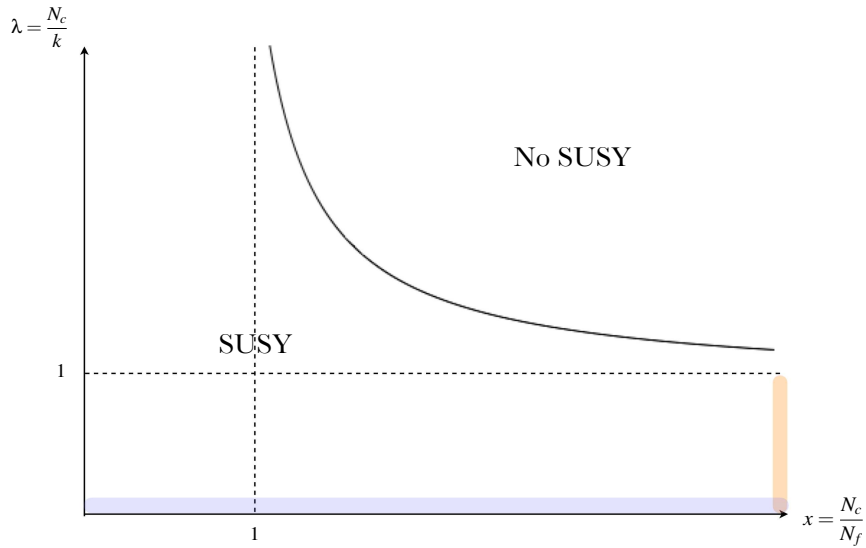}
\vspace{-.8cm}\bf
\caption{\it Phases of CS-SQCD in a $(\lambda,x)$ diagram. The blue region
at the bottom is the perturbative region of the electric theory. The orange region 
at the right corner is where the topological $\NN=2$ CS theory is recovered.}
\label{xlambda}
\end{figure}

Accordingly, the phases of CS-SQCD are plotted in Fig.\ \ref{xlambda}. There 
are two basic regions: a region where the vacuum is supersymmetric and a 
region where supersymmetry is spontaneously broken. The supersymmetric 
region has a corner at $x \to \infty$, $\lambda\in [0,1)$ and a corner at $x\in [0,1)$, 
$\lambda \to \infty$. The first corner, colored orange in Fig.\ \ref{xlambda}, is the 
place where $N_f \to 0$ and the theory becomes topological. At the second corner 
the coupling $\lambda$ is infinitely large.

It is interesting to note that the value $x=1$ is also special for the $SU(N_c)$ 
SQCD theory in four dimensions. The IR behavior of SQCD is characterized by
the single parameter $x$ and the supersymmetric vacuum is lifted  for $N_f<N_c$, 
$i.e.$ for $x>1$. In CS-SQCD we have instead two parameters characterizing
the theory, $\lambda$ and $x$. From the point of view of the stability of the 
supersymmetric vacuum, SQCD is similar to the infinitely strongly coupled 
regime of CS-SQCD where $\lambda\gg 1$. It is unclear whether this observation 
implies a deeper connection between three- and four-dimensional dynamics.

Finally, there are two regions where CS-SQCD admits a weakly coupled 
description. An obvious one is the region at the bottom, which is colored blue 
in Fig.\ \ref{xlambda}, where $\lambda \ll1$. The other one is the supersymmetric
region close to the critical SUSY breaking curve, $i.e.$ the region with 
$\lambda \lesssim \frac{x}{x-1}$, $x\geq 1$. Here, a weakly coupled description 
exists in terms of a Seiberg dual magnetic theory.

\subsection{Seiberg duality}
\label{subsec:SQCDduality}

It has been proposed that the $\NN=2$ CS-SQCD theories exhibit Seiberg 
duality \cite{Giveon:2008zn}. The dual magnetic theory is an $\NN=2$ CSM 
theory with gauge group $U(N_f+k-N_c)$ at level $k$. It has $N_f$ pairs of
quarks $q_i, \tilde q^i$ ($i=1,2,\cdots,N_f$) and a set of magnetic meson 
chiral superfields $M_j^i$ which are gauge singlets. As in four-dimensional
SQCD, the magnetic theory possesses a cubic superpotential
\beq
\label{SQCDdualityaa}
W_{mag}= M_j^i q_i \tilde q^j
~.
\eeq
In three dimensions, this is a classically relevant interaction. It generates an
RG flow and in the IR, where the magnetic theory is dual to the electric, this
interaction becomes marginal.

To obtain Seiberg duality in the context of the string theory configuration of Fig.\ 
\ref{CSSQCD} we displace sequentially the $N_f$ D5-branes and the $(1,k)$ 
bound state along the $x^6$ direction past the NS5-brane. When $N_f$ D5-branes
pass through the NS5-brane, $N_f$ D3-branes are created. Similarly, when
the $(1,k)$ bound state passes through the NS5-brane, $k$ D3-branes are
created. At the end of the process we obtain the configuration in Fig.\ 
\ref{CSSQCD}(b) whose low-energy dynamics is described by the magnetic 
theory presented above.

As a small parenthesis we note that in the topological limit $N_f \to 0$, 
Seiberg duality has a natural interpretation as level-rank duality in the bosonic 
$SU(N)_k$ WZW model \cite{Aharony:2008gk}. Indeed, at large $k$ we can 
integrate out the gluino field (whose mass is proportional to $k$) to obtain pure 
CS theory at the shifted level \cite{Kao:1995gf}
\beq
\label{SQCDdualityab}
k'=k-N_c
~.
\eeq
By the CS-WZW correspondence of \cite{Witten:1988hf} this theory is equivalent
to the (chiral) $SU(N_c)_{k-N_c}$ WZW model (for the moment we set the $U(1)$
part of the gauge group aside). Level-rank duality implies that this is equivalent to the 
$SU(k-N_c)_{N_c}$ WZW model, which, again by the CS-WZW correspondence, is 
equivalent to the $SU(k-N_c)$ pure CS theory at level $N_c$. Integrating in the gluinos 
(and putting back the $U(1)$ part) we recover the Seiberg dual $\NN=2$ CS 
theory with gauge group $U(k-N_c)$ and level $k$.

For $x>1$ Seiberg duality acts as a strong/weak coupling duality. Indeed, the 
magnetic 't Hooft coupling is
\beq
\label{SQCDdualityac}
\widetilde \lambda= \frac{N_f+k-N_c}{k}=1-\lambda \left(1-\frac{1}{x}\right)
~.
\eeq
Hence, when $\lambda\ll 1$, $\widetilde \lambda\sim 1$ and the magnetic
description is strongly coupled. Conversely, when $\widetilde \lambda \ll 1$, 
$\lambda \sim \left(1-x^{-1}\right)^{-1}>1$ and the electric theory is strongly 
coupled. This strong/weak relation disappears for $x<1$. In this case, 
both descriptions become strongly coupled simultaneously.

\subsection{Qualitative features of R-charges}
\label{subsec:SQCDRcharges}

We now come to one of the central questions in this paper -- how the R-charges
behave as we change the parameters of the CSM theory. In the electric version
of CS-SQCD the flavor symmetry $SU(N_f)\times SU(N_f)$ guarantees that all
the flavors $Q^i$, $\widetilde Q_i$ ($i=1,2,\cdots, N_f$) have the same $U(1)_R$
charge $R_Q=R_Q(N_c,N_f,k)$. In the large-$N$ limit $R_Q=R_Q(\lambda,x)$.
Similarly, in the magnetic theory all the flavors $q_i$, $\tilde q^i$ have the same
R-charge $R_q=R_q(\lambda, x)$. The magnetic theory also possesses the gauge
singlet elementary meson superfield $M^i_j$, whose R-charge we denote as
$R_M=R_M(\lambda, x)$.

The three functions $R_Q$, $R_q$ and $R_M$ are related by Seiberg duality.
The composite meson chiral superfields of the electric theory 
$\MM^i_j=Q^i \widetilde Q_j$ are mapped to the elementary fields $M^i_j$.
Hence,
\beq
\label{SQCDRchaa}
R_M=2R_Q
~.
\eeq
Moreover, in the IR of the magnetic theory the superpotential $W_{mag}$ 
\eqref{SQCDdualityaa} becomes marginal and
\beq
\label{SQCDchab}
R_M+2R_q=2
~.
\eeq
We conclude that there is one independent R-charge function, say $R_Q$, and
\beq
\label{SQCDchac}
R_q=1-R_Q~, ~ ~ R_M=2R_Q
~.
\eeq

In the four-dimensional $SU(N_c)$ SQCD we can determine $R_Q$ exactly
by demanding anomaly cancellation. The result is $R_Q=1-x$. In three dimensions 
there are no anomalies for continuous symmetries, hence we cannot proceed
in the same way. 

Alternatively, if we did not know about anomalies in four dimensions,
but we knew about Seiberg duality, it would still have been possible to determine 
$R_Q$ exactly in $\NN=1$ SQCD. By matching the baryon operators ($a_i$ are 
gauge indices here)
\beq
\label{SQCDchad}
B^{i_1\cdots i_{N_c}}=\epsilon^{a_1a_2\cdots a_{N_c}}Q_{a_1}^{i_1}\cdots 
Q_{a_{N_c}}^{i_{N_c}}~, ~ ~ 
\widetilde B^{i_1\cdots i_{N_c}}=\epsilon_{a_1a_2\cdots a_{N_c}}
\widetilde Q^{a_1}_{i_1}\cdots \widetilde Q^{a_{N_c}}_{i_{N_c}}
\eeq
to their magnetic duals one finds independently a new relation between 
$R_Q$ and $R_q$
\beq
\label{SQCDchae}
N_cR_Q=(N_f-N_c)R_q
\eeq
which determines $R_Q$ as above in agreement with the anomaly cancellation
condition. We cannot repeat this exercise in CS-SQCD, because the gauge group 
is $U(N_c)$ and there are no baryons to match. Hence, the $U(1)$ part of the gauge 
group, which is so crucial for the validity of Seiberg duality in CS-SQCD,\footnote{Indeed, 
if we simply dropped the $U(1)$ in CS-SQCD, baryon matching would have been 
problematic.} is also the reason why the exact form of the function $R_Q$ avoids a 
simple detection.

\begin{figure}[t!]
\centering
\includegraphics[width=13cm, height=8.3cm]{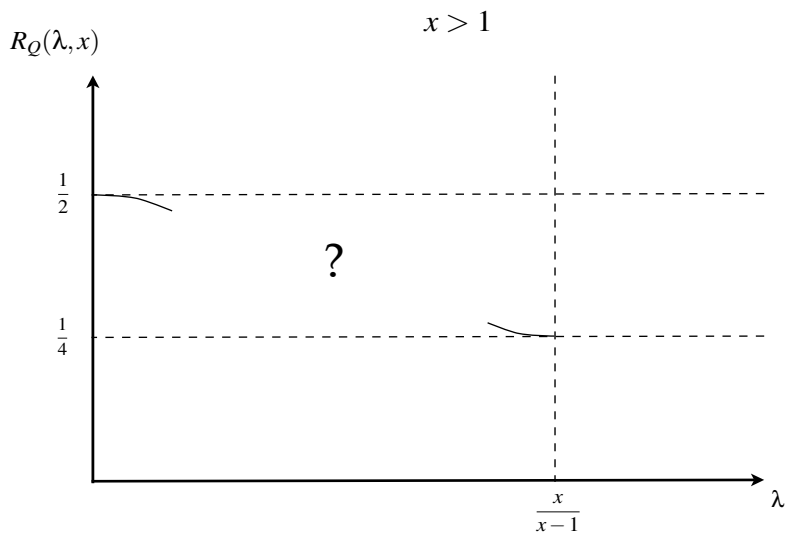}
\vspace{-.2cm}\bf
\caption{\it A plot of $R_Q$ as a function of $\lambda$ for fixed $x>1$. $R_Q$
interpolates between its classical value $\frac{1}{2}$ and the value $\frac{1}{4}$
at the critical point $\frac{x}{x-1}$ beyond which the theory exhibits spontaneous
breaking of supersymmetry. The exact behavior of the function in this interval is 
currently unknown.}
\label{RQfig}
\end{figure}

Some information about the behavior of $R_Q(\lambda, x)$ can be obtained in 
perturbative regimes. For $\lambda,\frac{1}{x} \ll 1$ a two-loop perturbative 
computation of $R_Q$ gives \cite{Gaiotto:2007qi} (we have taken the large-$N$ 
limit for simplicity)
\beq
\label{SQCDchaf}
R_Q(\lambda,x)=\frac{1}{2}-\frac{\lambda^2}{16}+\cdots
\eeq 
where the dots $\cdots$ indicate subleading corrections in $\lambda$ and
$\frac{1}{x}$. As usual, we observe that gauge interactions work to reduce the 
classical R-charge.

At strong coupling the behavior of the theory depends, as we said, on the value
of $x$. For $x<1$, the coupling $\lambda$ can grow to infinity and there is no
obvious regime which admits a weakly coupled description. For $x>1$, however,
Seiberg duality provides a weakly coupled description when $\lambda\lesssim 
\frac{x}{x-1}$. In this regime we can compute using the magnetic theory. At the
critical coupling $\lambda=\frac{x}{x-1}$, the rank of the dual gauge group
vanishes and the quark superfields disappear. The magnetic superpotential is 
absent and the superfields $M^i_j$ are free fields. This picture is supported by
the brane setup in Fig.\ \ref{CSSQCD}(b). At $\lambda=\frac{x}{x-1}$ the color
D3-branes are absent, there are no quark superfields from color-flavor
open strings and the IR dynamics is dominated by the free fields $M^i_j$
which arise from flavor-flavor open strings. Consequently, in this limit 
$R_M=\frac{1}{2}$ and $R_Q=\frac{1}{4}$. Moreover, at any value of 
$\lambda<\frac{x}{x-1}$ we must have $R_M\geq \frac{1}{2}$ by unitarity. 

Assuming $R_Q$ is a continuous function of $\lambda$
the emerging picture for $x>1$ is depicted in Fig.\ \ref{RQfig}. At the lower part 
of the interval $\left [0,\frac{x}{x-1}\right]$ $R_Q$ starts off at its classical value
$\frac{1}{2}$ and then decreases. At the upper end, $R_Q$ tends from above
to the value $\frac{1}{4}$ where the mesons saturate the unitarity 
bound.\footnote{The continuity of $R_Q$ at the critical coupling 
$\lambda=\frac{x}{x-1}$ is not immediately obvious from the magnetic 
theory point of view. Although a discontinuity is unlikely there in our opinion, 
a clear justification of this point would be desirable. A discontinuity of 
$R_Q$ at the critical coupling would imply that $R_Q$ tends to a finite
value between $\frac{1}{2}$ and $\frac{1}{4}$ as $\lambda \to \frac{x}{x-1}$,
but jumps discontinuously to the value $\frac{1}{4}$ at $\lambda=\frac{x}{x-1}$.
We proceed in the next section assuming this discontinuous behavior does not occur.} 
In this range $R_Q$ is necessarily greater than $\frac{1}{4}$, but how it behaves 
more precisely is currently unclear. It would be interesting, for example, to know if 
$R_Q(\lambda, x)$ is a monotonic function of $\lambda$ for fixed $x$ (this seems 
to be a natural expectation in the absence of superpotential interactions). It would 
also be interesting to know what happens when $x<1$.

\section{CS-SQCD theories with mesonic superpotentials}
\label{sec:SQCDsuperpot}

\vspace{-.3cm}
\subsection{Relevant mesonic superpotentials}
\label{subsec:SQCDrelevant}

We can deform the superconformal $\NN=2$ CS-SQCD theories by adding to 
the Lagrangian superpotential interactions that involve the chiral meson operators. 
For $x>1$ the qualitative picture of the previous section helps us understand which
are the relevant superpotential interactions that we can add.

The superpotential deformations
\beq
\label{SQCDrelaa} 
\delta W_1=m^j_iQ^i\widetilde Q_j~, ~ ~ 
\delta W_2=\frac{\alpha_2}{2} (Q^i \widetilde Q_j)(Q^j \widetilde Q_i)
\eeq
are relevant already at weak coupling $\lambda$. The first is a mass deformation,
the second is a deformation that drives the theory to an $\NN=3$ supersymmetry 
enhanced IR fixed point. This RG flow will be discussed in the next subsection.

Higher powers of the meson operators are classically irrelevant operators. We have
seen, however, that as we increase the coupling the R-charge $R_Q$ goes down 
achieving the minimum value $\frac{1}{4}$ at $\lambda=\frac{x}{x-1}$. Consequently, 
there is a critical value of $\lambda$ beyond which the sextic superpotential deformation 
\beq
\label{SQCDrelab}
\delta W_3=\alpha_3 (Q \widetilde Q)^3
\eeq
becomes relevant. We will discuss this deformation in subsection \ref{subsec:SQCDsextic}. 

The next (and last in the supersymmetric interval) power of mesons 
$(Q\widetilde Q)^4$ becomes marginal at the critical curve $\lambda=\frac{x}{x-1}$.
If Seiberg duality is to be trusted there, this point has a dual description in terms
of a WZ model for $M$ with superpotential
\beq
\label{SQCDrelac}
W_{mag}=\alpha_4 M^4
\eeq
The leading two-loop correction to the $\beta$-function of $\alpha_4$ is positive
(as follows quite generally from unitarity). Hence, at least perturbatively near 
$\alpha_4=0$, this perturbation is irrelevant and does not lead to a new fixed point.

\subsection{Quartic deformations}
\label{subsec:SQCDquartic}

The quartic deformation $\delta W_2$ in \eqref{SQCDrelaa} is classically
marginal, but quantum mechanically relevant. At small $\lambda$, the 
large-$N$ limit $\beta$-function for the coupling $\alpha_2$ is \cite{Gaiotto:2007qi}
\beq
\label{SQCDquaa}
\frac{d \alpha_2}{d t}=\frac{N_c^2}{(8\pi)^2}\alpha_2
\left[\alpha_2^2-\left(\frac{4\pi}{k}\right)^2\right]
~,
\eeq
where $t$ is the logarithm of the RG scale. Perturbing by $\delta W_2$ 
drives the theory to a new IR fixed point with $\alpha_2$ value 
\beq
\label{SQCDquab}
\alpha_{2,\NN=3}=\pm \frac{4\pi}{k}
\eeq
controlled by the CS level $k$. This is a fixed point with enhanced $\NN=3$ 
supersymmetry \cite{Gaiotto:2007qi} (see also below). Accordingly, the new 
R-symmetry is $SU(2)$. The $SU(N_f)\times SU(N_f)$ global flavor symmetry 
is broken down to the diagonal $SU(N_f)$ by the superpotential interaction. 

At the $\NN=3$ fixed point the superpotential interaction is again marginal 
and $R_Q$ has recovered its classical value $\frac{1}{2}$. The absence of
quantum corrections to $R_Q$ as we change $\lambda, x$ at this point is 
consistent with the fact that the $U(1)_R$ symmetry is now part of the larger
$SU(2)_R$ symmetry which precludes the presence of anomalous dimensions
for the meson operators $Q\widetilde Q$.

As a deformation of the $\NN=2$ CS-SQCD theory, the $\NN=3$ fixed points are 
still expected to exhibit Seiberg duality in terms of a $U(N_f+k-N_c)$ theory and 
spontaneous supersymmetry breaking at $N_c=N_f+k$. In what follows we will 
describe two different brane configurations in type IIB string theory that confirm 
these properties.

\subsubsection{Quartic couplings from brane configurations I}
\label{subsec:quarticI}

The first configuration of branes in type IIB string theory that we want to consider 
is similar to the configuration appearing in Fig.\ \ref{CSSQCD}(a). The only 
difference is a change in the orientation of the $(1,k)$ bound state in the (48)
and (59) planes. D3, D5, NS5 and $(1,k)$ branes are now oriented in the 
following way
\beq
\label{SQCDquba}
\begin{array}{r c l}
1~ NS5 ~&:&~~ 0~1~2~3~4~5
\\[1mm]
1~ (1,k)~&:&~~ 0~1~2~\left[ {3 \atop 7} \right]_\theta~ 
\left[ {4 \atop 8} \right]_\theta ~
\left[ {5 \atop 9} \right]_{-\theta}
\\[1mm]
N_c~ D3~&:&~~ 0~1~2~|6|
\\[1mm]
N_f~ D5~&:&~~ 0~1~2~7~8~9
\end{array}
\eeq
$\theta$ is still the $k$-controlled angle that appears in eq.\ \eqref{defSQCDac}.
This configuration expressly preserves $\NN=3$ supersymmetry in three dimensions
\cite{Kitao:1998mf}.

The low-energy description of this setup is in terms of a $U(N_c)$ $\NN=3$ CS
theory at level $k$ coupled to $N_f$ pairs of (anti)fundamentals $Q^i$, $\widetilde Q_i$
and a massive chiral superfield $X$ in the adjoint representation of the gauge group.
The Lagrangian includes the superpotential interactions
\beq
\label{SQCDqubb}
W_{\NN=3}=-\frac{k}{8\pi}\tr X^2+\widetilde Q_i X Q^i
~.
\eeq
At large $k$ the superfield $X$ can be integrated out to recover the quartic 
superpotential $\delta W_2$ with $\alpha_2=\frac{4\pi}{k}$.

Within string theory Seiberg duality follows, as in section \ref{sec:CSSQCD}, by 
moving the D5-branes and the $(1,k)$ bound state through the NS5-brane along 
$x^6$ to obtain a configuration of the form depicted in Fig.\ \ref{CSSQCD}(b). The 
dual gauge theory is a $U(N_f+k-N_c)$ $\NN=3$ CS theory at level $k$ with $N_f$ 
pairs of (anti)fundamentals $q_i$, $\tilde q^i$, a massive adjoint chiral superfield $X$
and the superpotential \eqref{SQCDqubb} with $Q, \widetilde Q$ replaced by
$q, \tilde q$. From the $s$-rule of brane dynamics we deduce that there is no
supersymmetric vacuum for $N_c>N_f+k$. 

Seiberg duality in this case acts in a self-similar way. A generalization 
to theories with superpotentials of the form \eqref{SQCDqubb}, but arbitrary power 
for $X$ ($\tr X^{n+1}$), will be considered in section \ref{subsec:hatoMspecial}.

\subsubsection{Quartic couplings from brane configurations II}
\label{subsec:quarticII}

The generic deformation by $\delta W_2$ can be obtained in string theory
within the following type IIB configuration 
\beq	
\label{SQCDquca}
\begin{array}{r c l}
1~ NS5 ~&:&~~ 0~1~2~3~4~5
\\[1mm]
1~ (1,k)~&:&~~ 0~1~2~\left[ {3 \atop 7} \right]_\theta~ 8 ~9
\\[1mm]
N_c~ D3~&:&~~ 0~1~2~|6|
\\[1mm]
N_f~ D5~&:&~~ 0~1~2~7~\left[ {4 \atop 8} \right]_\psi ~ 
\left[ {5\atop 9} \right]_\psi
\end{array}
\eeq
The rotation of the D5-branes by an arbitrary angle $\psi$ in the (48), (59)
planes provides the quartic coupling $\delta W_2$. Similar type IIA configurations 
related to four-dimensional $\NN=1$ SQCD theories with quartic superpotential
have been discussed in \cite{Giveon:2007ew} (an earlier discussion oriented also
towards three-dimensional $\NN=2$ gauge theories can be found in \cite{Aharony:1997ju}). 
The presence of the quartic coupling in this setup will be justified in a moment. 
$\theta$ is again given by eq.\ \eqref{defSQCDac}. At the special value $\psi=\frac{\pi}{2}$ 
the D5-branes are oriented along 012789 and we reproduce the configuration that gives 
the $\NN=2$ CS-SQCD theory without superpotential interactions.

For generic angle $\psi$ the configuration \eqref{SQCDquca} preserves $\NN=2$ 
supersymmetry in three dimensions. A quick verification of this fact appears in 
appendix \ref{app:HWsusies}. The low-energy field theory exhibits $\NN=3$ 
supersymmetry enhancement for a special value of the quartic coupling, and 
therefore a special value of the angle $\psi$ (see below). This effect is not 
visible in the brane configuration.

Before discussing further the low-energy gauge theory description of this setup
it will be convenient to move the D5 and $(1,k)$ fivebranes along the $x^6$ 
direction past the NS5-brane. Once again, this motion leads to a Seiberg dual
configuration of the form depicted in Fig.\ \ref{CSSQCD}(b). At low energies
the dynamics of this configuration is described by the magnetic version of 
$\NN=2$ CS-SQCD (level $k$ and gauge group $U(N_f+k-N_c)$) with a 
mass deformed superpotential
\beq
\label{SQCDqucc}
\widetilde W=\sqrt{\mu} M^i_j q_i \tilde q^j + \frac{\alpha}{2} M_i^j M_j^i
~.
\eeq
$\mu$ is a scale with the dimension of mass which was kept implicit before.
The extra mass term for the meson $M$ appears because of the rotation of
the D5-branes. It captures the fact that the $N_f$ D3-branes stretching between 
the D5-branes and the $(1,k)$ bound state can no longer move freely in the (89) 
plane. The mass parameter $\alpha$ is related to the rotation angle $\psi$ via 
the relation
\beq
\label{SQCDqucd}
\alpha=\mu \cot \psi
~.
\eeq
By integrating out the massive fields $M^i_j$ we obtain a superpotential with
a quartic interaction for the magnetic quarks
\beq
\label{SQCDquce}
\widetilde W=-\frac{\tan \psi}{2}(q\tilde q)^2
~.
\eeq

Returning to the electric description, we recognize that the deformation which is
dual to $M^2$ is $(Q\widetilde Q)^2$. Hence, in the presence of the rotated 
D5-branes the electric theory includes the quartic superpotential interaction 
\beq
\label{SQCDqucf}
W=\frac{\beta^2 \cot\psi}{2} (Q\widetilde Q)^2
~,
\eeq
where $\beta$ is the proportionality constant that appears in the duality relation
$M^i_j=\frac{\beta}{\sqrt\mu} Q^i \widetilde Q_j$. Requiring that the $\NN=3$ 
supersymmetry enhancement occurs simultaneously in the electric and magnetic 
theories gives
\beq
\label{SQCD3susy}
\tan\psi_{\NN=3}=\frac{4\pi}{k}~, ~ ~ \beta=\pm \frac{4\pi}{k}
~.
\eeq

Again, we observe that Seiberg duality exchanges two versions of the same 
theory -- the rank of the gauge group is dualized and the quartic 
superpotential coupling is essentially inversed.

\subsection{Sextic deformations}
\label{subsec:SQCDsextic}

The sextic operator $(Q\tilde Q)^3$ has classical dimension $\Delta_6=3$,
and is therefore irrelevant at small coupling $\lambda$. As we increase the 
coupling for $x>1$ the scaling dimension $\Delta_6=6R_Q(\lambda, x)$ goes 
down (presumably monotonically) until it reaches the minimum value 
$\frac{3}{2}$ at the supersymmetry breaking boundary $\frac{x}{x-1}$. This 
qualitative picture predicts that there is a critical coupling $\lambda^*$ where 
the sextic operator becomes marginal. For this coupling
\beq
\label{SQCDsexaa}
R_Q(\lambda^*,x)=\frac{1}{3}~, ~ ~ x>1
~.
\eeq

When the coupling lies in the range $\lambda^*<\lambda \leq \frac{x}{x-1}$ we
can add this operator to the electric theory Lagrangian to generate an RG flow 
towards a new IR fixed point. Near the critical coupling $\lambda^*$ the 
deformation $\delta W_3$ (see eq.\ \eqref{SQCDrelab}) is slightly relevant 
and an IR fixed point is expected to exist at perturbative values of $\alpha_3$. 
It is nevertheless difficult to compute the $\beta$-function in conformal perturbation
theory in this case since the theory is at finite coupling $\lambda$. 

Applying Seiberg duality to the deformed theory we obtain a magnetic
dual at level $k$ with gauge group $U(N_f+k-N_c)$ and superpotential 
\beq
\label{SQCDsexab}
\widetilde W=\sqrt \mu \, M q\tilde  q+\tilde \alpha_3 M^3
~.
\eeq
At weak magnetic coupling $\widetilde \lambda=1-\lambda\frac{x-1}{x}\ll  1$
the operator $M^3$ is relevant ($\Delta (M^3)\sim \frac{3}{2}<2$) and the 
$\beta$-function is controlled to leading order in $\widetilde \lambda$ by a WZ
model for the field $M$ with cubic superpotential. This $\beta$-function is expected
to have a zero at a finite value of $\tilde \alpha_3$.

\section{CS-SQCD theories with one adjoint chiral superfield}
\label{sec:1adj}

\vspace{-.3cm}
\subsection{Brief review of known results}
\label{subsec:1adjrev}

In this section we will discuss the properties of a more complex set of theories.
These are $U(N_c)$ $\NN=2$ CSM theories at Chern-Simons level $k$ coupled to
$N_f$ pairs of (anti)fundamental chiral superfields $Q^i$, $\widetilde Q_i$ and one 
superfield $X$ in the adjoint representation of the gauge group. In the absence of 
superpotential interactions we will call this theory the $\hat {\sf A}$ theory following a 
notation applied to analogous four-dimensional gauge theories in \cite{Intriligator:2003mi}.

Applying the general arguments of \cite{Gaiotto:2007qi}, we deduce that the $\widehat {\sf A}$ 
theory is superconformal. Moreover, we will present a picture suggesting that, unlike 
the CS-SQCD theory without the adjoint superfield, in the $\widehat {\sf A}$ theory there is no 
range of parameters where supersymmetry is spontaneously broken. The global symmetry is
$SU(N_f)\times SU(N_f)\times U(1)_a\times U(1)_X\times U(1)_R$.

The classical chiral ring consists of the single trace operators
\beq
\label{1adjrevaa}
\tr X^{n+1}~, ~ ~ n=0,1,\cdots
~,
\eeq
made out of the adjoint chiral superfield $X$, and the generalized meson operators
\beq
\label{1adjrevab}
Q^i X^n \widetilde Q_j~, ~ ~ n=0,1,\cdots
~.
\eeq
Since we consider theories with gauge group $U(N_c)$ there are no baryon operators.
Gauge invariant baryon-like operators constructed out of 't Hooft monopole operators
\cite{'tHooft:1977hy} can exist though. We will not, however, consider such operators 
in this paper.

The addition of the superpotential interaction 
\beq
\label{1adjrevaca}
W_{n+1}=\frac{g_0}{n+1} \tr X^{n+1}
\eeq
to the Lagrangian truncates the above chiral ring to the finite subset
\beq
\label{1adjrevac}
\tr X^\ell~, ~ ~ Q^i\widetilde Q_j~, ~ ~ Q^iX^\ell \widetilde Q_j~, ~ ~ \ell=1,\cdots, n-1
~.
\eeq
We will call the resulting 1-adjoint CS-SQCD theories ${\sf A}_{n+1}$. The global 
symmetry of these theories is $SU(N_f)\times SU(N_f)\times U(1)_a \times U(1)_R$. 
The $U(1)_X$ symmetry has been broken explicitly by the superpotential $W_{n+1}$.

Some properties of the ${\sf A}_{n+1}$ theories, like Seiberg duality and spontaneous 
breaking of supersymmetry, can be deduced easily from a brane construction in string theory 
\cite{Niarchos:2008jb}. The relevant construction is a simple generalization of the type IIB 
string theory setup appearing in Fig.\ \ref{CSSQCD}. It involves 
\beq
\label{1adjrevac}
\begin{array}{r c l}
n~ NS5 ~&:&~~ 0~1~2~3~4~5
\\[1mm]
1~ (1,k)~&:&~~ 0~1~2~\left[ {3 \atop 7} \right]_\theta~ 8~9
\\[1mm]
N_c~ D3~&:&~~ 0~1~2~|6|
\\[1mm]
N_f~ D5~&:&~~ 0~1~2~7~8~9
\end{array}
\eeq
The new superpotential parameter $n$ is encoded in the number of NS5-branes 
along 012345. For $n=1$ the added superpotential is quadratic in the superfield $X$. 
Integrating out $X$ we recover the CS-SQCD theories of section \ref{sec:CSSQCD}.

The r\^ole of the superpotential interaction $\tr X^{n+1}$ in this setup can be identified
in the following way. Displacing the $n$ NS5-branes in the (89) plane to $n$ 
different points $a_\ell=x_\ell^8+ix_\ell^9$ $(\ell=1,2,\cdots,n)$ forces the $N_c$ 
D3-branes to break up into $n$ groups of $r_1$ D3-branes ending on the $a_1$ 
positioned NS5-brane, $r_2$ D3-branes ending on the $a_2$ positioned NS5-brane 
$etc.$ with
\beq
\label{1adjrevad}
\sum_{\ell=1}^n r_\ell=N_c
~.
\eeq
The new configuration describes vacua of the gauge theory where the diagonal
matrix elements of the complex scalar in the superfield $X$ acquire expectation 
values $a_i$. These vacua are captured in gauge theory by a polynomial 
superpotential interaction of the form
\beq
\label{1adjrevae}
W(X)=\sum_{\ell=0}^n \frac{g_\ell}{n+1-\ell}X^{n+1-\ell}
~.
\eeq
For generic coefficients $\{ g_\ell \}$ the superpotential has $n$ distinct minima
$\{ a_\ell \}$ related to $\{ g_\ell \}$ via the relation
\beq
\label{1adjrevaf}
W'(x)=\sum_{\ell=0}^n g_\ell x^{n-\ell}=g_0 \prod_{\ell=1}^n (x-a_\ell)
~.
\eeq
In gauge theory the partition integers $r_\ell$ label the number of eigenvalues
of the $N_c\times N_c$ matrix $X$ residing in the $\ell$-th minimum of the potential
$V=|W'(x)|^2$. When all the expectation values are distinct the adjoint field is
massive and the gauge group is Higgsed
\beq
\label{1adjrevag}
U(N_c)\to U(r_1)\times U(r_2)\times \cdots \times U(r_n)
~.
\eeq
In this vacuum we obtain $n$ decoupled copies of the $\NN=2$ CS-SQCD theories
at level $k$ with $N_f$ flavor multiplets.

The condition for the existence of a supersymmetric vacuum in the 
$U(r_\ell)$ $\NN=2$ CS-SQCD theory is
\beq
\label{1adjrevai}
r_\ell \leq k+N_f~, ~  ~ 1\leq \ell \leq n
~.
\eeq
Summing over $\ell$ and sending all the $a_i$'s to zero we obtain a 
condition for the existence of a supersymmetric vacuum in the ${\sf A}_{n+1}$
theory
\beq
\label{1adjrevaj}
N_c \leq n(k+N_f)
~.
\eeq
The same condition can be obtained in string theory with the use of the $s$-rule of 
brane dynamics \cite{Niarchos:2008jb}.

A Seiberg dual version of the ${\sf A}_{n+1}$ theory can be obtained by
moving the D5 and $(1,k)$ fivebranes past the $n$ NS5-branes along the $x^6$ 
direction as in Fig.\ \ref{CSSQCD}(b). This duality, which was the subject of Ref.\
\cite{Niarchos:2008jb}, is a three-dimensional CS analog of Kutasov duality
in four-dimensional $\NN=1$ gauge theory \cite{Kutasov:1995ve}. The magnetic
${\sf A}_{n+1}$ theory is an $\NN=2$ CSM theory at CS level $k$ and gauge group
$U(n(N_f+k)-N_c)$ coupled to $N_f$ pairs of chiral multiplets $q_i$, $\tilde q^i$, an 
adjoint chiral superfield $Y$ and $n$ magnetic mesons $M_\ell$ ($\ell=1,\cdots, n$),
each of which is an $N_f\times N_f$ matrix. There is a dual tree-level superpotential
\beq
\label{1adjrevak}
\widetilde W_{n+1}=-\frac{g_0}{n+1}\tr Y^{n+1}+\sum_{\ell=1}^n M_\ell \tilde q Y^{n-\ell}q
~.
\eeq

All these statements have important consequences for the structure of the $\hat{\sf A}$ and
${\sf A}_{n+1}$ theories which we now proceed to uncover.

\subsection{New results on R-charges}
\label{subsec:1adjRcharges}

In the $\widehat {\sf A}$ theory the flavor symmetry $SU(N_f)\times SU(N_f)$
guarantees that all the flavors $Q^i$, $\widetilde Q_i$ have the same $U(1)_R$
charge $R_Q=R_Q(N_c,N_f,k)$. The $U(1)_R$ charge of the superfield $X$ is 
another function $R_X=R_X(N_c,N_f,k)$. We will not be able to say much about
the function $R_Q$ in this paper, but there is a number of interesting statements
that we can make about the function $R_X$ using information about the properties
of the ${\sf A}_{n+1}$ theories. 

The classical value of $R_X$ in the $\widehat {\sf A}$ theory is $\frac{1}{2}$. Hence, 
at weak coupling, $\lambda=\frac{N_c}{k}\ll 1$, the chiral operators $\tr X^2$, $\tr X^3$ 
are relevant, the chiral operator $\tr X^4$ is classically marginal and the chiral operators 
$\tr X^n$ ($n>4$) are irrelevant. Therefore, adding $\tr X^n$ ($n>4)$ to the Lagrangian at 
weak coupling will not modify the IR behavior of the theory.

The fact that supersymmetry can be spontaneously broken in the ${\sf A}_{n+1}$
theory proves that as we increase the coupling $\lambda$, $R_X$ receives in the 
$\widehat{\sf A}$ theory large negative anomalous contributions which eventually 
make the operator $\tr X^{n+1}$ relevant. At the supersymmetry breaking value of 
$\lambda$ the operator $\tr X^{n+1}$ modifies the IR behavior of the theory so 
drastically that the supersymmetric vacuum is lifted.\footnote{A similar observation 
relating the anomalous dimensions of $X$ and vacuum stability can also be found 
in Refs.\ \cite{Kutasov:1995np,Kutasov:2003iy} in the context of four-dimensional 
$\NN=1$ adjoint SQCD theories. I thank David Kutasov for a discussion that prompted 
me to think more about this relation.} Presicely how $R_X$ decreases and whether or not 
an operator $\tr X^{n+1}$ can become relevant in the $\widehat{\sf A}$ theory depends 
on the value of $x$, $i.e.$ the ratio $N_c/N_f$.

In terms of the parameters $\lambda$ and $x$ a supersymmetric vacuum in the 
${\sf A}_{n+1}$ theory exists for all $\lambda$ if $x\leq n$, and for $\lambda$ such that
\beq
\label{1adjRchaa}
\lambda \leq \lambda_{n+1}^{\rm SUSY}=\frac{nx}{x-n}
\eeq
for $x>n$. 

Now let us fix the value of $x$ and discuss what happens in the $\hat{\sf A}$
theory as we increase the coupling $\lambda$. At small $\lambda$ and large $x$ 
a perturbative calculation based on the results of \cite{Gaiotto:2007qi} gives
\beq
\label{1adjRchab}
R_X(\lambda, x)\sim \frac{1}{2}-\left(2+\frac{4}{3x}\right)\lambda^2+\cdots
~,
\eeq
where the dots $\cdots$ denote subleading contributions.
As we further increase $\lambda$ the R-charge $R_X$ continues to decrease. For all
integers $n\leq [x]$ ($[x]$ denotes the integer part of $x$) there is a sequence of critical
values 
\beq
\label{1adjRchac}
0=\lambda^*_2=\lambda^*_3=\lambda^*_4<\lambda^*_5<\cdots
<\lambda^*_n<\lambda^*_{n+1}<\cdots<\lambda^*_{[x]}<\lambda^*_{[x]+1}
\eeq
where each time one of the chiral operators $\tr X^{n+1}$ ($n \leq [x]$) becomes marginal.
By definition, $\lambda^*_{n+1}$ is the point where the chiral operator $\tr X^{n+1}$ 
becomes marginal, $i.e.$ the point where
\beq
\label{1adjRchad}
R_X(\lambda^*_{n+1},x)=\frac{2}{n+1}
~.
\eeq
Above $\lambda^*_{n+1}$ the operator $\tr X^{n+1}$ is relevant. Adding it to the Lagrangian
in this range of parameters drives the theory to a new fixed point -- the ${\sf A}_{n+1}$ theory. 
By further increasing the coupling in the ${\sf A}_{n+1}$ theory we reach the SUSY
breaking point $\lambda_{n+1}^{\rm SUSY}$ where the supersymmetric vacuum is 
destabilized. This observation provides an upper bound on the exact value of 
$\lambda^*_{n+1}$
\beq
\label{1adjRchae}
\lambda^*_{n+1}<\lambda_{n+1}^{\rm SUSY}=\frac{nx}{x-n}
~.
\eeq

\begin{figure}[t!]
\centering
\includegraphics[width=14cm, height=9.5cm]{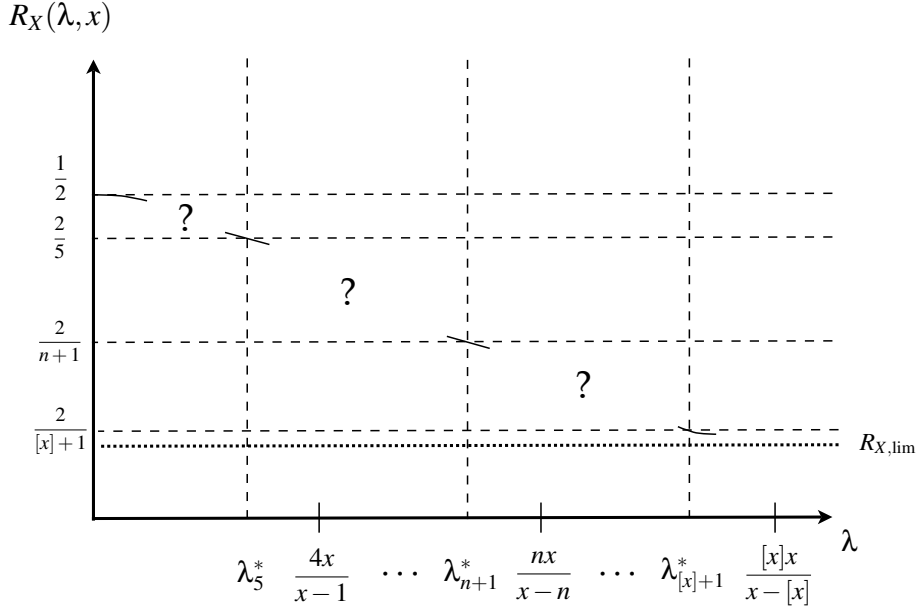}
\vspace{-.8cm}\bf
\caption{\it A plot of $R_X$ in the $\hat{\sf A}$ theory as a function of $\lambda$ for fixed 
$x$. $R_X$ interpolates between its classical value $\frac{1}{2}$ and a limiting value 
$R_{X,\rm lim}$ in the range $\frac{1}{2([x]+2)}<R_{X,\rm lim}<\frac{2}{[x]+1}$. $\lambda^*_{n+1}$ 
is the point where the chiral operator $\tr X^{n+1}$ becomes marginal. The critical point 
$\frac{nx}{x-n}$ is where the ${\sf A}_{n+1}$ theory exhibits spontaneous breaking of 
supersymmetry.}
\label{RXfig}
\end{figure}

The emerging picture is depicted graphically in Fig.\ \ref{RXfig}. The R-charge $R_X$
decreases monotonically as we increase $\lambda$ making more and more single
trace chiral operators $\tr X^{n+1}$ relevant. Beyond the critical coupling $\lambda^*_{[x]+1}$,
$R_X$ approaches a limiting lowest value $R_{X,\rm lim}>0$, which lies somewhere 
inside the interval $\left( \frac{1}{2([x]+2)},\frac{2}{[x]+1}\right)$. The lower bound of this 
interval arises in the following way. 

The scaling dimension of the operator $\tr X^{[x]+2}$
is $\Delta_{[x]+2}=([x]+2)R_X(\lambda, x)$. We cannot exclude the possibility that this 
operator becomes relevant beyond some coupling, but even if it does it cannot destabilize 
the supersymmetric vacuum in the ${\sf A}_{[x]+2}$ theory. If the function $R_X(\lambda, x)$ 
continues to decrease monotonically towards zero, a value of $\lambda$ will be reached
eventually where $\Delta_{[x]+2}=\frac{1}{2}$. Beyond this point the operator $\tr X^{[x]+2}$ 
becomes a free field and decouples from the rest of the theory. Hence, in this regime, a 
deformation by a superpotential interaction linear in $\tr X^{[x]+2}$ will break the supersymmetry,
something that we know from the above analysis cannot happen. We conclude that 
$\Delta_{[x]+2}>\frac{1}{2}$ for all $\lambda$, which implies $R_{X,\rm lim}>\frac{1}{2([x]+2)}$.

A qualitatively similar situation occurs in the four-dimensional analog of this theory -- the IR 
of the $\NN=1$ adjoint SQCD theory -- as we vary the single parameter $x$. In that case, we can 
compute exactly where the critical values of $x$ lie using $a$-maximization \cite{Kutasov:2003iy}.

In our case, it would be nice to know how fast an operator $\tr X^{n+1}$ becomes relevant.
In other words, it would be nice to have an estimate of the magnitude of the difference 
$\lambda_{n+1}^{\rm SUSY}-\lambda_{n+1}^*$. Fortunately, such an estimate is within 
the power of our current considerations. 

We have observed that as we increase $\lambda$ the scaling dimension of the generic 
chiral operator $\tr X^{n+1}$, $\Delta_{n+1}=(n+1)R_X(\lambda, x)$, decreases.
If $\Delta_{n+1}$ reaches the unitarity bound $\frac{1}{2}$ at some value of $\lambda$,
call it $\lambda_{n+1}^{\rm max}$, then beyond this point the operator $\tr X^{n+1}$
becomes free and decouples from the rest of the theory. For reasons similar to the ones 
outlined above this cannot happen before we reach the SUSY breaking point 
$\lambda_{n+1}^{\rm SUSY}$ of the ${\sf A}_{n+1}$ theory. We can determine 
$\lambda^{\rm max}_{n+1}$ in the following way.

As we increase $\lambda$ beyond $\lambda^*_{n+1}$, we reach the critical coupling 
$\lambda_{n'+1}^*$ ($n'>n$) of another chiral operator $\tr X^{n'+1}$. There is an
integer $n'$ for which $\tr X^{n'+1}$ is marginal and simultaneously $\tr X^{n+1}$
becomes free. This occurs when
\beq
\label{1adjRchaf}
\Delta_{n+1}=(n+1)R_X(\lambda^*_{n'+1},x)=\frac{2(n+1)}{n'+1}=\frac{1}{2}
~~ \Leftrightarrow ~~ n'=4n+3
~.
\eeq
This, of course, will be true as long as $n'\leq [x]$, $i.e.$ $n\leq \frac{[x]-3}{4}$. Assuming
this inequality, we deduce that 
\beq
\label{1adjRchag}
\lambda^{\rm max}_{n+1}=\lambda^*_{n'+1}=\lambda^*_{4(n+1)}
\eeq
and our previous observations imply 
\beq
\label{1adjRchai}
\lambda_{n+1}^*<\frac{nx}{x-n}<\lambda^{\rm max}_{n+1}=\lambda^*_{4(n+1)}
~.
\eeq
The second inequality provides a lower bound to $\lambda^*_{n+1}$
\beq
\label{1adjRchaj}
\frac{\left[\frac{n-3}{4}\right] x}{x-\left[\frac{n-3}{4}\right]}<\lambda^*_{n+1}
\eeq
and gives an estimate to the difference $\lambda^{\rm SUSY}_{n+1}-\lambda^*_{n+1}$
provided $n\leq \frac{[x]-3}{4}$. At large values of $x$, $i.e.$ when $N_f\ll N_c$, the 
combination of the lower and upper bounds \eqref{1adjRchaj} and \eqref{1adjRchae} 
gives the inequalities
\beq
\label{1adjRchak}
\left[\frac{n-3}{4}\right]<\lambda^*_{n+1}<n
~.
\eeq

\subsection{More on Seiberg duality}
\label{hatoASeiberg}

Let us denote compactly as $\widetilde {\widehat{\sf A}}$ the set of $U(N_c)$ $\NN=2$ 
CSM theories at CS level $k$ without superpotential interactions that are coupled to 
$N_f$ quark multiplets $q_i$, $\tilde q^i$, an adjoint chiral superfield $Y$ and $n$ gauge 
singlet superfields $M_\ell$ ($\ell=1,2,\cdots,n$), each of which is an $N_f\times N_f$
matrix. The magnetic description of the $U(N_c)$ ${\sf A}_{n+1}$ theory arises from the
$U(n(N_f+k)-N_c)$ $\widetilde {\widehat{\sf A}}$ theory after the superpotential 
interaction \eqref{1adjrevak} is added to the Lagrangian.

The $U(n(N_f+k)-N_c)$ $\widetilde {\widehat{\sf A}}$ theories are superconformal field 
theories with large-$N$ parameters
\beq
\label{hatoSeibaa}
\widetilde \lambda=\frac{n(N_f+k)-N_c}{k}=n-\lambda\left(1-\frac{n}{x}\right)
~, ~ ~ 
\widetilde x=\frac{n(N_f+k)-N_c}{N_f}=n+x\left(\frac{n}{\lambda}-1\right)
~.
\eeq
They are weakly coupled when $\widetilde \lambda\ll 1$. Assuming $n>3$, all the 
operators appearing in $\widetilde W_{n+1}$ (eq.\ \eqref{1adjrevak}) are irrelevant
in the perturbative regime, except for the cubic operator $M_n \tilde q q$ 
and the quartic $M_{n-1}\tilde q Yq$. Both of them are relevant (the quartic operator 
is classically marginal with perturbatively negative anomalous dimension). Adding 
the operators $M_n \tilde q q$ and $M_{n-1}\tilde q Yq$ to the Lagrangian as 
superpotential interactions drives the theory to a new interacting fixed point.

More and more terms in the superpotential \eqref{1adjrevak} are expected to become
relevant in the $\widetilde {\widehat{\sf A}}$ theory as we increase $\widetilde \lambda$.
Notice that the elementary and composite mesons ($M_\ell$ and $\tilde q Y^{n-\ell} q$
respectively) are Legendre-transform conjugate variables in the magnetic theory. 
Therefore, depending on whether the term $M_\ell \tilde q Y^{n-\ell} q$ is relevant or
not in the magnetic superpotential, we should include either $M_\ell$ or $\tilde q Y^{n-\ell} q$ 
in the spectrum of independent operators.

Ultimately, as we increase $\widetilde \lambda$ we should encounter a critical coupling 
$\widetilde \lambda^*_{n+1}$ above which the operator $\tr Y^{n+1}$ is relevant and 
both the electric and magnetic theories flow towards the ${\sf A}_{n+1}$ fixed point. 
We can write this critical coupling as
\beq
\label{hatoSeibab}
\widetilde \lambda^*_{n+1}=n-\lambda_{n+1}^{**}\left(1-\frac{n}{x}\right)
\eeq
In terms of the electric 't Hooft coupling the magnetic theory enters the phase with
$\widetilde \lambda>\widetilde \lambda^*_{n+1}$ when
\beq
\label{hatoSeibac}
\lambda\left( -1+\frac{n}{x}\right)>\lambda_{n+1}^{**}\left(-1+\frac{n}{x}\right)
~.
\eeq
Demanding that the ${\sf A}_{n+1}$ fixed point can be obtained simultaneously by 
adding the relevant operator $\tr X^{n+1}$ to the electric theory means
$x>n$ and $\lambda>\lambda_{n+1}^*$. All these conditions can be met 
if and only if
\beq
\label{hatoSeibad}
\lambda^*_{n+1}<\lambda_{n+1}^{**}<\lambda_{n+1}^{\rm SUSY}
~.
\eeq
Verifying this prediction requires a strong analytic tool -- the analog of 
$a$-maximization in four dimensions.

Inside the `window' $[\lambda^*_{n+1},\lambda^{**}_{n+1}]$ 
both the electric and magnetic theories flow to the same IR fixed point where
\beq
\label{hatoSeibae}
R_X=R_Y=\frac{2}{n+1}
~.
\eeq 
From the remaining R-charges ($R_Q$ for $Q$, $\widetilde Q$, 
$R_q$ for $q$, $\tilde q$, and $R_{M_\ell}$ for $M_\ell$) only one is 
independent. The map between electric and magnetic meson fields and 
the marginality of the mesonic superpotential interactions implies the relations
\beq
\label{hatoSeibaf}
R_Q+R_q=\frac{2}{n+1}~, ~ ~ R_{M_\ell}=\frac{2(\ell-1)}{n+1}+2R_Q~, ~ ~ 
\ell=1,2,\cdots, n
~.
\eeq

Finally, matching the chiral rings and mapping superpotential deformations 
of the electric ${\sf A}_{n+1}$ theory to its magnetic dual is something that
can be achieved precisely as in four dimensions \cite{Kutasov:1995ss}.
The classical chiral rings do not match, but the quantum chiral rings do. 
Since the adjoint field $X$ in the electric theory is an $N_c\times N_c$ matrix
it obeys automatically, by virtue of the Caley-Hamilton theorem, the restrictions 
that follow from the characteristic equation
\beq
\label{hatoSeibag}
f(X)=0~, ~ ~ {\rm where}~ ~f(z)\equiv \det(z-X)
~.
\eeq  
To obtain the quantum chiral ring, the classical chiral ring relations must 
be supplemented by the characteristic equation of the magnetic theory.
Analogous statements apply to the magnetic theory. 

Mapping superpotential deformations of the form \eqref{1adjrevae} under 
Seiberg duality entails the steps taken in the four-dimensional adjoint SQCD 
theories in \cite{Kutasov:1995ss}. A minor difference with the analysis of 
\cite{Kutasov:1995ss} arises from the fact that here we discuss $U(N_c)$, 
instead of $SU(N_c)$, gauge groups.

\subsection{A special case and comments on holography}
\label{subsec:1adjpure}

Many supersymmetric CSM theories, with prototype the $\NN=6$ CSM 
theories in \cite{Aharony:2008ug}, admit a holographic dual description 
in terms of either string theory or M-theory on some $AdS_4$ background
of the form $AdS_4\times \MM$, with $\MM$ being some compact manifold.
One may wonder whether the $\NN=2$ CSM theories in this section 
have a similar holographic $AdS_4$ description.

A special case without the usual complications of fundamental matter is
the case of the $U(N_c)$ 1-adjoint CS-SQCD theories with $N_f=0$.
Besides the superconformal fixed points $\widehat {\sf A}$, labeled by the
integers $N_c$, $k$, new fixed points ${\sf A}_{n+1}$ can be obtained by 
adding the superpotential interactions
\beq
\label{1adjpureaa}
W_{n+1}=\frac{g_0}{n+1}\tr X^{n+1}
\eeq
for any integer $n\geq 1$ and $\lambda$ greater than a critical value 
$\lambda_{n+1}^*$. Each of the $U(N_c)$ ${\sf A}_{n+1}$ theories admits a 
Seiberg dual description in terms of another ${\sf A}_{n+1}$ theory at the same 
level $k$ but different gauge group $U(nk-N_c)$. The dual description disappears 
when the supersymmetric vacuum is spontaneously broken in the original theory. 
The condition for the existence of a supersymmetric vacuum in the $U(N_c)$ theory is 
\beq
\label{1adjpureab}
N_c \leq nk~ ~ \Leftrightarrow ~~ \lambda\leq n
~.
\eeq
There is a regime, $\lambda\in [\lambda_{n+1}^{**},n]$ in the notation of the 
previous subsection, where the operator $\tr X^{n+1}$ is an irrelevant operator
in the dual description. In that case, Seiberg duality exchanges an
${\sf A}_{n+1}$ fixed point with an $\widehat{\sf A}$ fixed point.

The standard large-$N$ reasoning suggests that these superconformal field 
theories have a holographic string theory description. Symmetries imply 
that this description involves non-critical strings on some $AdS_4\times S^1$
background, presumably with curvature of order the string scale. The symmetries
of $AdS_4$ reproduce the field theory superconformal symmetries and the 
$S^1$ the internal $U(1)_R$ symmetry. 

The possibility of a holographic description for the $\widehat {\sf A}$ fixed points 
was also discussed in \cite{Gaiotto:2007qi}. The setup in \cite{Gaiotto:2007qi}
involves $N_c$ M5-branes wrapping a special Lagrangian Lens space $S^3/\Z_k$
in a Calabi-Yau three-fold. There is no $AdS_4$ solution for this system in 
supergravity which implies that $\alpha'$ corrections are indeed important.

The ${\sf A}_{n+1}$ theories are also related to wrapped M5-branes. For example, 
we can realize the ${\sf A}_2$ theory with finite coupling $g_0=-\frac{k}{4\pi}$ 
as a special case of the configuration \eqref{SQCDquba} with $N_f=0$. This
setup is a special case of the configurations analyzed in Ref.\ \cite{Aharony:2008gk}.
Compactifying the $x^6$ direction, T-dualizing, and lifting to M-theory 
converts the suspended D3-branes into `fractional M2-branes', $i.e.$ 
M5-branes wrapping a vanishing 3-cycle at a $\Z_k$ orbifold point.

\section{Two-adjoint theories: RG flows from the $\widehat {\sf O}$ theory}
\label{sec:hato}

So far we have discussed $\NN=2$ CSM theories with an arbitrary number of 
fundamental/anti-fundamental pairs of chiral multiplets and one adjoint chiral
multiplet. These theories comprise a small set in the larger domain of $\NN=2$
theories with two, instead of one, chiral superfields. In this section, we will explore
RG flows and fixed points in this wider setup. The zoo of $\NN=1$ superconformal
field theories with two adjoint chiral superfields in four dimensions was discussed,
using $a$-maximization techniques, in Ref.\ \cite{Intriligator:2003mi}. In that work 
an intriguing ADE classification of fixed points was observed. We will find that a subset
of our three-dimensional superconformal field theories admits a similar classification.

\subsection{Relevant deformations to $\widehat{\sf A}$, $\widehat{\sf D}$, $\widehat{\sf E}$}
\label{subsec:hatoADE}

Our starting point is a theory which, mimicking \cite{Intriligator:2003mi}, we will
call the $\widehat {\sf O}$ theory. By definition, this theory is a $U(N_c)$ $\NN=2$ 
CSM theory at level $k$ coupled to $N_f$ pairs of fundamental/anti-fundamental
chiral superfields and two chiral superfields in the adjoint representation. We will
denote the adjoint chiral superfields $X$ and $X'$. The $\widehat {\sf O}$ theory
has no superpotential interactions. It is superconformal by the arguments
of Ref.\ \cite{Gaiotto:2007qi} and possesses the global symmetry group
$SU(N_f)\times SU(N_f)\times SU(2) \times U(1)_X \times U(1)_R$. The 
$SU(2)$ symmetry rotates the $(X,X')$ doublet. Because of this symmetry
the $U(1)_R$ charges of $X$ and $X'$ are identical and will be denoted by 
$R_X$, which is a function of the parameters $k, N_c, N_f$.

The chiral ring includes single trace operators of the form $\tr X^n {X'}^{n'}$ 
$(n,n'=0,1,\cdots)$ up to arbitrary permutations of the fields $X$, $X'$ inside
the trace. The scaling dimension of such operators is 
\beq
\label{hatoADEaa}
(n+n')R_X(N_c,N_f,k)
~.
\eeq
Hence, any of the $(n,n')$ operators is relevant when
\beq
\label{hatoADEab}
R_X(N_c,N_f,k)<\frac{2}{n+n'}
~.
\eeq
In that case, we can add the operator to the $\widehat {\sf O}$ Lagrangian as 
a superpotential interaction to generate an RG flow towards a new IR fixed 
point.

As in the analysis of the previous sections we expect the function $R_X$ to
decrease as $\lambda$ becomes larger and larger and the gauge interactions
stronger. This can be verified explicitly with a two-loop calculation in the 
perturbative regime \cite{Gaiotto:2007qi} 
\beq
\label{hatoADEac}
R_X(\lambda,x)\sim \frac{1}{2}-\left(3+\frac{4}{3x}\right)\lambda^2+\cdots
~.
\eeq
Since we have very limited information about the non-perturbative 
behavior of the function $R_X$, we will concentrate, in what follows, to operators 
that are either already classically relevant or classically marginal but quantum 
mechanically relevant.

Deformations involving only the operators $\tr X$, $\tr X'$ will not be considered since
they lead to F-term equations that cannot be solved. We will consider the
following (inequivalent) quadratic and cubic superpotential deformations:
\begin{itemize}
\item[(1)] $W=\tr XX'$. In this case, both chiral superfields $X$, $X'$ are massive
and the RG flow interpolates between the $\widehat{\sf O}$ theory and CS-SQCD
with $N_f$ flavors.
\item[(2)] $W=\tr {X'}^2$. The chiral superfield $X'$ is massive and can be integrated 
out. The IR fixed point is the $\widehat{\sf A}$ theory that was analyzed in the previous 
section.
\item[(3)] $W=\tr X{X'}^2$. The RG flow leads to a new IR fixed point, which we will
call the $\widehat {\sf D}$ theory.
\item[(4)] $W=\tr {X'}^3$. The IR fixed point arising from this RG flow will be named 
$\widehat {\sf E}$.
\end{itemize}

The (inequivalent) quartic deformations $W=\tr X^4, \tr X^3 X', \tr X^2 {X'}^2,
\tr XX'XX'$ are marginal at $\lambda=0$ but relevant at $\lambda>0$. The
generated RG flows are a special case of the flows studied in perturbation 
theory in \cite{Gaiotto:2007qi}. We will not have anything new to add concerning
this case. Instead, we will proceed to examine RG flows away from the
$\widehat{\sf A}$, $\widehat{\sf D}$, $\widehat{\sf E}$ theories which have 
certain similarities with RG flows in two-dimensional $\NN=(2,2)$ 
Landau-Ginzburg models and four-dimensional $\NN=1$ SQCD theories
with two adjoints.

\subsection{Flows from $\widehat{\sf A} \to {\sf A}_{n+1}$}
\label{subsec:hatoA}

RG flows from the $\widehat{\sf A}$ theory to ${\sf A}_{n+1}$ fixed
points occur when the superpotential $W=\tr X^{n+1}$ is relevant. The 
conditions for this to happen were explored in the previous section. 
We have not shown explicitly that the IR of this RG flow is indeed a fixed point 
of the $\beta$-function. This is a hard question because the ${\sf A}_{n+1}$ 
theory is non-perturbative for generic $n$. The only exception is the case $n=3$, 
$i.e.$ the case of a quartic deformation. Here one can show the existence of a 
perturbative fixed point with a two-loop computation of the $\beta$-function 
\cite{Gaiotto:2007qi}.

\subsection{Flows from $\widehat{\sf D} \to {\sf D}_{n+2}$}
\label{subsec:hatoD}

The $\widehat{\sf D}$ theory arises from the $\widehat{\sf O}$ theory
after the deformation by the superpotential interaction 
$W_{\widehat{\sf D}}=\tr X{X'}^2$. In this theory the chiral ring of gauge 
invariant operators is subject to the relations coming from the 
$W_{\widehat{\sf D}}$ equations of motion
\beq
\label{hatoDaa}
\d_{X'}W_{\widehat{\sf D}}=\{ X, X'\}=0~, ~~
\d_XW_{\widehat{\sf D}}={X'}^2=0
~.
\eeq
The first equation is particularly convenient, because we can use it to freely
re-order the fields $X$, $X'$ inside traces (up to a minus sign). Using these
equations we find that the chiral ring is generated by the single trace operators
\beq
\label{hatoDab}
\tr X^\ell~, ~ ~ \ell\geq 1~, ~ ~ \tr X'
\eeq
and the meson operators
\beq
\label{hatoDac}
M_{\ell,s}=\tilde Q X^\ell {X'}^s Q~, ~ ~ \ell\geq 0~, s=0,1
~.
\eeq
Note that $\tr X^n {X'}=0$ because of the first equation in \eqref{hatoDaa} and 
the cyclicity of the trace.

In the $\widehat{\sf D}$ theory the superpotential interaction $W_{\widehat{\sf D}}$
is marginal and imposes the following constraint on the R-charges
\beq
\label{hatoDad}
R_X+2R_{X'}=2
~.
\eeq
Hence, the independent R-charges that need to be determined as functions
of the parameters $k,N_c,N_f$ are the R-charge $R_X$ of $X$ and the common
R-charge $R_Q$ of the (anti)fundamental multiplets $Q$, $\widetilde Q$.

Any of the chiral operators \eqref{hatoDab}, \eqref{hatoDac} can be used
to deform the Lagrangian of the $\widehat{\sf D}$ theory. We will focus on 
superpotential deformations involving the first set \eqref{hatoDab}. Without
prior knowledge of the R-charges $R_X$, $R_{X'}$ it is unclear which of 
these deformations are relevant and if so for what range of parameters. 
If this case is similar to the $\widehat {\sf A}$ theory we might expect that $R_X$
decreases as we increase $\lambda$ and, because of eq.\ \eqref{hatoDad},
at the same time $R_{X'}$ increases. Assuming this is true, and that
there is a range of parameters where the operator $\tr X^{n+1}$ is relevant,
we can deform by the superpotential interaction $\Delta W=\tr X^{n+1}$ to flow 
towards a tentative new fixed point which we will call ${\sf D}_{n+2}$. In what 
follows, we will argue in favor of these flows and will propose that the ${\sf D}_{n+2}$
fixed points exist and exhibit non-trivial properties, among them Seiberg duality.

Before proceeding further, notice that the $n=1$ deformation involves the
superpotential
\beq
\label{hatoDae}
W_{n=1}=\frac{g}{2}\tr X^2+a \tr X{X'}^2
~.
\eeq
The field $X$ is massive in this case and by integrating it out we get the low energy
superpotential
\beq
\label{hatoDaf}
W=-\frac{a^2}{2g}\tr {X'}^4
~.
\eeq
In this way, we recover the 1-adjoint CS-SQCD fixed point ${\sf A}_4$.

\subsubsection{Stability bounds and their consequences}
\label{subsec:hatoDstability}

The crucial element that allowed us in the $\widehat{\sf A}$ theories to determine 
the qualitative behavior of the R-charge $R_X$ was a bound on $\lambda$ for 
the stability of the supersymmetric vacuum in the ${\sf A}_{n+1}$ theories.
We could read these bounds directly from the $s$-rule in a string theory setup, or
by deforming slightly the superpotential (as in eq.\ \eqref{1adjrevae}), flowing to 
a product of CS-SQCD vacua and then using the condition for the existence of
a supersymmetric vacuum in CS-SQCD. We can repeat the second argument 
in the ${\sf D}_{n+2}$ field theories. A similar analysis was performed in 
Ref.\ \cite{Intriligator:2003mi} for the four-dimensional ${\sf D}_{n+2}$ theories.
Since the argument is identical in our case we will focus on the main points 
highlighting elements particular to our case and defer the reader to 
\cite{Intriligator:2003mi} for additional details.

The core of the argument centers around the precise way in which the chiral
ring truncates in the presence of the ${\sf D}_{n+2}$ superpotential
\beq
\label{hatoDstaa}
W_{{\sf D}_{n+2}}=\tr X^{n+1}+\tr X{X'}^2
~.
\eeq
For simplicity, we keep the superpotential coefficients in front of each term 
implicit. The relations coming from this superpotential are
\beq
\label{hatoDstab}
\{X,X'\}=0~, ~ ~ X^n+{X'}^2=0
~.
\eeq
What happens to the chiral ring depends crucially on whether $n$ is odd 
or even.

For odd $n$ there is a drastic truncation of the chiral ring. Using the relations
\eqref{hatoDstab} one can show that ${X'}^3=0$. The classical chiral ring 
includes the single trace operators
\beq
\label{hatoDstac}
\tr X^{\ell-1}~, ~ \ell=1,\cdots, n~, ~~ \tr X'~, ~ ~ \tr {X'}^2~, ~ ~ 
\tr X^{2m}{X'}^2~, ~ m=1,\cdots,\frac{n-1}{2}
~,
\eeq
where the order of $X$ and $X'$ does not matter because of the first relation 
in \eqref{hatoDstab}, and the $3nN_f^2$ mesons
\beq
\label{hatoDstad}
\MM_{\ell,s}=\widetilde Q X^{\ell-1} {X'}^{s-1}Q~, ~ ~ \ell=1,\cdots, n~, ~ ~
s=1,2,3
~.
\eeq

To determine the conditions for the existence of a supersymmetric vacuum
we deform the superpotential $W_{{\sf D}_{n+2}}$ by lower order terms
\beq
\label{hatoDstae}
W=\tr \left( F_{n+1}(X)+X{X'}^2+X'\right)
\eeq
where $F_{n+1}(X)$ is a degree $n+1$ polynomial in $X$. The F-term 
equations for this superpotential are
\beq
\label{hatoDstaf}
\{X,X'\}=-1~, ~ ~ {X'}^2+\d_X F_{n+1}(X)=0
~.
\eeq
The vacua of the field theory are solutions of these equations. The irreducible
representations of the algebra defined by \eqref{hatoDstaf} are $n+2$ different
one-dimensional representations and $\frac{n-1}{2}$ two-dimensional 
representations \cite{Cachazo:2001gh,Intriligator:2003mi}. The general vacuum 
has $r_a$ copies of the $a$-th one-dimensional representation $(a=1,\cdots,n+2)$,
and $s_b$ copies of the $b$-th two-dimensional representation $(b=1,\cdots,\frac{n-1}{2})$.
In such a vacuum the gauge group is Higgsed to
\beq
\label{hatoDstag}
U(N_c)\to \prod_{a=1}^{n+2} U(r_a) \prod_{b=1}^{\frac{n-1}{2}} U(s_b)
~ ~ ~ {\rm with} ~~~ 
\sum_{a=1}^{n+2} r_a+\sum_{b=1}^{\frac{n-1}{2}}2 s_b=N_c
\eeq
and both adjoint chiral superfields are massive. Each $U(r_a)$ factor has
$N_f$ pairs of flavor multiplets and each $U(s_b)$ factor $2N_f$ flavor pairs 
\cite{Intriligator:2003mi}. Hence, each factor is a CS-SQCD theory with some number
of flavor multiplets ($N_f$ or $2N_f$). The condition for the existence of a 
supersymmetric vacuum in CS-SQCD (see eq.\ \eqref{defSQCDad})
implies for the product \eqref{hatoDstag}
\beq
\label{hatoDstai}
r_a \leq N_f+k~, ~ ~ s_b\leq 2(N_f+k)~, ~ ~ a=1,\cdots, n+2~, ~ ~ b=1,\cdots,\frac{n-1}{2}
~.
\eeq
Summing up these inequalities we obtain a condition for the existence of a
supersymmetric vacuum in the ${\sf D}_{n+2}$ theory
\beq
\label{hatoDstaj}
N_c \leq 3 n(N_f+k)
~.
\eeq
Notice that for $n=1$ this formula reproduces the condition \eqref{1adjrevaj} for the
${\sf A}_4$ theory which is consistent with the observation in eq.\ \eqref{hatoDaf}
that ${\sf D}_3$ is essentially the ${\sf A}_4$ theory.

For even $n$ the situation is more complex. In this case, the classical chiral ring 
does not truncate and a bound like \eqref{hatoDstaj} cannot be derived classically.
However, a bound may exist at the quantum level. The fact that we can add a 
superpotential deformation $\Delta W=\tr X^{n'+1}$ to the ${\sf D}_{n+2}$ theory with 
$n'<n$, $n$ odd and $n'$ even, to flow from the ${\sf D}_{n+2}$ theory to the 
${\sf D}_{n'+2}$ theory suggests that the even $n$ theories also have a stability 
bound. This is based on the natural expectation that by adding a more relevant 
term to the superpotential it will be easier for the vacuum to get destabilized. 
A natural hypothesis is that the stability bound for $n$ even continues to obey 
the same form that was found in eq.\ \eqref{hatoDstaj}. Further motivation for this
hypothesis will be provided in a moment. In four-dimensional $\NN=1$ adjoint 
SQCD theories of the ${\sf D}_{n+2}$ type this assumption is corroborated by 
the $a$-conjecture \cite{Intriligator:2003mi}.

Assuming the validity of \eqref{hatoDstaj} for all $n$ as a working hypothesis,
we can deduce a qualitative picture for the $\lambda$-dependence of the 
R-charge $R_X$ in the $\widehat{\sf D}$ theory, which is similar to that in the 
1-adjoint $\widehat{\sf A}$ theory. In terms of the 't Hooft parameters $\lambda,x$
a supersymmetric vacuum exists in the ${\sf D}_{n+2}$ theory for all $\lambda$
if $x\leq 3n$, and for $\lambda$ such that
\beq
\label{hatoDstak}
\lambda\leq \lambda^{\rm SUSY}_{n+2}=\frac{3nx}{x-3n}
\eeq
for $x>3n$.
Hence, as we increase $\lambda$ in the $\widehat {\sf D}$ theory the R-charge 
$R_X$ decreases and for $n<\frac{x}{3}$ the operator $\tr X^{n+1}$ becomes 
marginal at a critical coupling 
\beq
\label{hatoDstal}
\lambda^*_{n+2}<\lambda^{\rm SUSY}_{n+2}=\frac{3nx}{x-3n}
~.
\eeq
The qualitative behavior of $R_X$ is the same as in Fig.\ \ref{RXfig}
with the obvious modifications. The analog of the inequalities \eqref{1adjRchak} is
\beq
\label{hatoDstam}
3 \left[ \frac{n-3}{4} \right] <\lambda^*_{n+2}<3n
~.
\eeq
The behavior of $R_{X'}$ is fixed in terms of the relation \eqref{hatoDad}.

\subsubsection{Seiberg-Brodie duality}
\label{subsec:hatoDduality}

It has been argued \cite{Brodie:1996vx} that the four-dimensional 
${\sf D}_{n+2}$ theories exhibit Seiberg duality. It is tempting to propose 
that the ${\sf D}_{n+2}$ CSM theories in this section also exhibit Seiberg 
duality augmenting the known list of $\NN=2$ Chern-Simons theories 
with Seiberg duals. 

Previous experience with Seiberg duality in $\NN=2$ CSM theories
\cite{Giveon:2008zn,Niarchos:2008jb} shows that the rank of the dual
gauge group encodes the spontaneous supersymmetry breaking boundary 
of the original theory in a simple fashion. Extending this feature to the ${\sf D}_{n+2}$
theories we propose that they have a dual magnetic description in terms of
an $\NN=2$ CSM theory at the same level $k$ and gauge group 
$U(3n(N_f+k)-N_c)$.\footnote{A dual rank of this universal form, independent
of whether $n$ is even or odd, is further motivation for the postulated extension 
of the inequality \eqref{hatoDstaj} to $n$ even.} The matter content and the superpotential 
interactions of the dual magnetic theory are partially fixed by the chiral ring structure.
The details work as in four dimensions, hence we propose that the matter
content of the dual theory includes $N_f$ pairs of (anti)fundamental multiplets
$q_i$, $\tilde q^i$, two adjoint chiral superfields $Y,Y'$ and $3nN_f^2$ 
gauge singlets $(M_{\ell,s})^i_j$ ($\ell=1,\cdots, n$, $s=1,2,3$, $i,j=1,\cdots, N_f$),
which are the magnetic duals of the meson operators \eqref{hatoDstad}.
The dual tree-level superpotential is
\beq
\label{hatoDduaa}
\widetilde W_{{\sf D}_{n+2}}=
\tr Y^{n+1}+\tr Y{Y'}^2+\sum_{\ell=1}^n \sum_{s=1}^3 
M_{\ell,s} \tilde q Y^{n-\ell} {Y'}^{3-s}q
~.
\eeq
Weak evidence for the validity of this duality is presented in appendix \ref{app:EviBrodie}.

When $3n<x$ this duality works as a strong/weak duality. The magnetic
't Hooft coupling $\widetilde \lambda$ is related to the electric coupling $\lambda$
in the following way
\beq
\label{hatoDduab}
\widetilde \lambda=3n-\left(1-\frac{3n}{x}\right) \lambda
~.
\eeq
Repeating the discussion of section \ref{hatoASeiberg} we anticipate 
a window $[\lambda^*_{n+2},\lambda^{**}_{n+2}]$ inside which both
$\tr X^{n+1}$ and its dual $\tr Y^{n+1}$ are relevant operators.

\subsection{Flows from $\widehat{\sf E}$}
\label{subsec:hatoE}

The $\widehat{\sf E}$ theory arises from the $\widehat{\sf O}$ theory after
the superpotential deformation
\beq
\label{hatoEaa}
W_{\widehat{\sf E}}=\tr {X'}^3
~.
\eeq 
The equations of motion for this superpotential impose the classical chiral 
ring relation ${X'}^2=0$ and truncate the chiral ring of the $\widehat {\sf O}$
theory to the operators
\beq
\label{hatoEab}
\tr X_{i_1} \cdots X_{i_n}\, , ~ ~ \widetilde Q Q\, , ~ ~ 
\widetilde Q X_{i_1}\cdots X_{i_n} Q\, , ~ ~ {\rm with}~~
i_1,i_2\cdots =1,2~, ~ ~ X_1=X\, , \, X_2=X'~, ~ ~ n=1,2,\cdots
\eeq 
with the provision that there are no adjacent $X'$ operators in the above 
combinations (including adjacency via cyclic permutation).

In the $\widehat {\sf E}$ theory the R-charge of the $X'$ field is fixed by
the superpotential
\beq
\label{hatoEac}
R_{X'}=\frac{2}{3}
~.
\eeq
What remains to be computed are the common R-charge $R_Q$ of the 
quarks $Q$, $\widetilde Q$, and the R-charge $R_X$ of the adjoint field $X$.

Since we lack an exact analytic tool that allows us to compute these charges 
we will restrict our attention to the weak coupling regime and deformations
that involve only the single trace operators of the adjoint fields.
To leading order in $\lambda$, $R_X$ has the classical value $\frac{1}{2}$.
Hence, any superpotential deformation of the form 
\beq
\label{hatoEad}
\Delta W=\tr X_{i_1} \cdots X_{i_N}~, ~ ~ N=n+n'
\eeq
with $n$ insertions of $X$ and $n'$ insertions of $X'$ will be relevant as long as
\beq
\label{hatoEae}
\frac{2n'}{3}+\frac{n}{2}<2 ~\Leftrightarrow~ 4n'+3n<12
~.
\eeq
This inequality allows several possibilities.

The linear deformation by $\tr X$ gives F-term equations that cannot be solved,
hence it is discarded. The linear deformation by $\tr X'$ is, however, allowed.

There are three quadratic deformations giving rise to the following RG flows
\begin{subequations}
\begin{eqnarray}
\label{hatoEaf}
&\Delta W=\tr {X'}^2~, ~ ~ &\widehat{\sf E}\to \widehat{\sf A}
~,\\
\label{hatoEag}
&\Delta W=\tr X^2~, ~ ~ &\widehat{\sf E}\to {\sf A}_3
~,\\
\label{hatoEai}
&\Delta W=\tr XX'~, ~ ~ &\widehat{\sf E}\to {\rm CS-SQCD}
~.
\end{eqnarray}
\end{subequations}

Finally, the inequality \eqref{hatoEae} allows the cubic deformation
\beq
\label{hatoEaj}
\Delta W=\tr X^2 X'~, ~ ~ \widehat{\sf E}\to {\sf D}_4
~.
\eeq
The deformation $\Delta W=\tr X^3$ is equivalent to $\tr X^2 X'$ via a 
change of variables.

It is natural to expect that as we increase the coupling $\lambda$ the
R-charge $R_X$ will receive more and more negative contributions from
the gauge interactions allowing for higher degree relevant superpotential 
deformations. It is impossible, however, to determine if and when this 
happens without more detailed information. In this respect, it is worth 
pointing out that in four dimensions the only (independent) higher degree 
deformations (becoming relevant at some range of parameters) are 
\cite{Intriligator:2003mi}
\begin{subequations}
\bea
\label{hatoEak}
&&{\sf E}_6~:~ \Delta W=\tr X^4
~,\\
&&{\sf E}_7~:~ \Delta W=\tr X^3 X'
~,\\
&&{\sf E}_8~:~ \Delta W=\tr X^5
~.
\eea
\end{subequations}
It is an interesting problem to determine if there is a similar pattern 
in our CSM theories in three dimensions.

\subsection{Comments on mesonic deformations}
\label{subsec:hatoM}

Our list of RG flows from the $\widehat{\sf O}$ theory above is certainly 
not exhaustive and admits more possibilities. Another large class of RG 
flows is generated by superpotential deformations involving mesonic 
operators. 

An example is provided by the superpotential deformation
\beq
\label{hatoMaa}
\Delta W=\widetilde Q_i X Q^i
\eeq
in the $\widehat {\sf O}$ theory. Classically this cubic deformation is relevant
and leads to a new fixed point which, again following the four-dimensional
nomenclature of \cite{Intriligator:2003mi}, we will call $\widehat{\sf O}_{\rm M}$.
Further deformations of this theory by single trace chiral operators are possible
and will be discussed in the next section.

Another possibility, which is visible at weak coupling, involves the quartic
meson operators $\widetilde Q X^2 Q$ and $\widetilde Q XX' Q$. These operators
are classically marginal, but receive negative anomalous dimensions and 
generate flows that can be described in perturbation theory as in Ref.\ 
\cite{Gaiotto:2007qi}.

In general, deformations by higher order mesonic operators, $e.g.$ 
$\widetilde Q X^\ell Q$, may be possible, but precise knowledge of whether and 
when these operators can become relevant depends on information about the 
R-charge $R_Q$ for which we have not been able to say much in this paper.

\section{RG flows from the $\widehat {\sf O}_{\rm M}$ theory}
\label{sec:hatom}

\vspace{-.3cm}
\subsection{Input from a Hanany-Witten setup}
\label{subsec:hatoMHW}

It is instructive to consider a straightforward generalization of the brane 
configuration appearing in Fig.\ \ref{CSSQCD} where instead of one
NS5-brane and one $(1,k)$ fivebrane bound state we consider $n$
NS5-branes and $n'$ $(1,k)$ fivebrane bound states $(n,n'=1,2,\cdots)$.
To summarize the configuration we have
\beq
\begin{array}{r c l}
\label{hatoMHWaa}
n~ NS5 ~&:&~~ 0~1~2~3~4~5
\\[1mm]
n'~ (1,k)~&:&~~ 0~1~2~\left[ {3 \atop 7} \right]_\theta~ 8~9
\\[1mm]
N_c~ D3~&:&~~ 0~1~2~|6|
\\[1mm]
N_f~ D5~&:&~~ 0~1~2~7~8~9
\end{array}
\eeq
with the angle $\theta$ given by eq.\ \eqref{defSQCDac}.

The low energy effective field theory that describes the dynamics of this
system is a $U(N_c)$ $\NN=2$ CSM theory at level $k$ coupled to $N_f$
(anti)fundamentals $Q^i$, $\widetilde Q_i$ and two adjoint chiral superfields
$X$, $X'$. As in section \ref{sec:1adj}, the fields $X$, $X'$ are present to 
describe fluctuations of the D3-branes along the $(89)$ and the $(45)$ planes
respectively. There is also a non-trivial superpotential \cite{Elitzur:1997hc}
\beq
\label{hatoMHWab}
W_{n,n'}=\frac{g_0}{n+1}\tr X^{n+1}+\frac{g'_0}{n'+1}\tr {X'}^{n'+1}+
\sum_{i=1}^{N_f} m \widetilde Q_i X' Q^i
~.
\eeq
The third mesonic superpotential interaction encodes an important difference
between the $X$ and $X'$ fields. When we displace the $n'$ $(1,k)$ bound
states along the $(45)$ plane, leaving the D5-branes fixed, we make the quark
multiplets $Q$ and $\widetilde Q$ massive with the same mass of order 
$\langle X' \rangle$. This effect is accounted for by the Yukawa superpotential 
coupling $m$. Its presence breaks the global $SU(N_f)\times SU(N_f)$
flavor symmetry to a diagonal $SU(N_f)$.

The analysis of the vacuum structure of this configuration suggests that the
matrices $X$, $X'$ can be diagonalized independently \cite{Elitzur:1997hc}.
Ref.\ \cite{Giveon:1998sr} proposed that the superpotential $W_{n,n'}$
includes an additional quartic coupling $\tr [X,X']^2$ for which we will have
little to say here.

From the brane configuration and the $s$-rule of brane dynamics we read
off the following condition for the existence of a supersymmetric vacuum
\beq
\label{hatoMHWac}
N_c\leq n N_f+nn'k
~.
\eeq
As a trivial check, for $n'=1$ the field $X'$ is massive. In that case, it can be 
integrated out to obtain the ${\sf A}_{n+1}$ theory and eq.\ \eqref{hatoMHWac} 
reproduces eq.\ \eqref{1adjrevaj}. 

Another interesting piece of information that we obtain from this brane 
configuration is Seiberg duality in a large class of 2-adjoint CS-SQCD 
theories. By moving the $N_f$ D5-branes and the $n'$ $(1,k)$ bound 
states along the $x^6$ direction past the $n$ NS5-branes we obtain a 
configuration similar to the one appearing in Fig.\ \ref{CSSQCD}(b). This 
configuration comprises now of $nN_f$ flavor D3-branes and 
$n(N_f+n'k)-N_c$ color D3-branes and realizes a dual 
magnetic description of the original theory. This description is provided 
by an $\NN=2$ CSM theory with gauge group $U(n(N_f+n'k)-N_c)$, 
CS level $k$ and the following matter content: $N_f$ quark pairs $q_i, \tilde q^i$, 
two adjoint chiral superfields $Y,Y'$ and $n$ magnetic meson fields $M_\ell$
($\ell=1,\cdots, n$), each of which is an $N_f\times N_f$ matrix. There
is a non-trivial superpotential
\beq
\label{hatoMHWad}
\widetilde W_{n,n'}=\frac{\tilde g_0}{n+1}\tr Y^{n+1}
+\frac{{{\tilde g}'}_0}{n'+1}\tr {Y'}^{n'+1}
+\sum_{i=1}^{N_f} \widetilde m \tilde q Y' q
+\sum_{\ell=1}^n M_\ell \tilde q Y^{n-\ell} q
\eeq
with a possible $\tr [Y,Y']^2$ term as in the electric theory. A four-dimensional 
analog of this duality was formulated in \cite{Elitzur:1997hc}.

The theories appearing in this context can be regarded as deformations
of the 2-adjoint $\widehat {\sf O}_{\rm M}$ CSM theories defined in 
the previous section. In what follows we will explore some of the 
consequences that the above statements have for their dynamics.

\subsection{Special case: $n=1$, $n'\geq 1$}
\label{subsec:hatoMspecial}

We mentioned that the special case $n\geq 1$, $n'=1$ reduces (by integrating out 
the massive $X'$ field) to the ${\sf A}_{n+1}$ theories which were analyzed before. 
Another interesting special case is the one with $n=1$ and $n'>1$. In this
case the superfield $X$ is massive and can be integrated out. Then, one is
left with an 1-adjoint CS-SQCD theory with superpotential
\beq
\label{hatoMspaa}
W_{n'+1}=\frac{g'_0}{n'+1}\tr {X'}^{n'+1}+\sum_{i=1}^{N_f}m_i \widetilde Q_i X' Q^i
~.
\eeq
In the brane construction all $m_i$ are equal.

We can view this theory as another deformation of the $\widehat {\sf A}$ theory. 
Adding the classically relevant superpotential interaction $\widetilde Q X' Q$ to 
the Lagrangian we flow towards a new set of IR fixed points, which we will
call collectively $\widehat {\sf A}_{\rm M}$. Then, we deform further to 
a new set of theories ${\sf A}_{{\rm M},n'+1}$ by adding the superpotential
interactions $\tr {X'}^{n'+1}$. Notice that the special case $n'=1$ with 
$g'_0=-\frac{k}{4\pi}$ reproduces the theory \eqref{SQCDqubb} which flows
to an $\NN=3$ fixed point with a quartic superpotential for the quarks.

The $s$-rule derived condition for the existence of a supersymmetric vacuum
in the ${\sf A}_{{\rm M},n'+1}$ theories is
\beq
\label{hatoMspab}
N_c\leq N_f+n'k
~.
\eeq
For $N_c>N_f$, $i.e.$ $x>1$, it implies that as we increase the coupling $\lambda$ 
in the $\widehat {\sf A}_{\rm M}$ theory the R-charge of $X'$, $R_{X'}$, decreases 
while more and more single trace operators $\tr {X'}^{n'+1}$ are becoming sequentially 
relevant. The value of the coupling where $\tr {X'}^{n'+1}$ becomes marginal is
\beq
\label{hatoMspac}
\lambda^*_{n'+1}<\lambda^{\rm SUSY}_{n'+1}=\frac{n'x}{x-1}
~.
\eeq
A picture similar to the one depicted in Fig.\ \ref{RXfig} for the ${\sf A}_{n+1}$ theory
is emerging with an important difference. Assuming $x>1$, a spontaneous 
breaking of supersymmetry occurs in the present case for arbitrarily large $n'$. 
Therefore, $R_{X',\rm lim}=0$ and any operator $\tr {X'}^{n'+1}$ can become relevant 
as long as we make the coupling $\lambda$ large enough. Furthermore, when all 
$m_i$ are equal the R-charges of the quarks $Q^i$, $\widetilde Q_i$ are also equal 
and can be denoted by a single function $R_Q$. In the $\widehat {\sf A}_{\rm M}$ 
theory the mesonic superpotential interaction is marginal, hence there is a simple 
relation between
$R_Q$ and $R_{X'}$
\beq
\label{hatoMspad}
R_Q=1-\frac{1}{2}R_{X'}
~.
\eeq

Seiberg duality relates this theory to a $U(N_f+n'k-N_c)$ magnetic version with 
a single adjoint chiral superfield $Y'$ and $N_f$ pairs of quarks $q_i$, $\tilde q^i$. 
The magnetic superpotential is
\beq
\label{hatoMspae}
\widetilde W_{n'+1}=\frac{g'_0}{n'+1} \tr {Y'}^{n'+1}+\sum_{i=1}^{N_f} \widetilde m_i
\tilde q^i Y' q_i
~.
\eeq
Comparing with the dual superpotential \eqref{1adjrevak} we observe that the
$n'$ elementary meson superfields $M_\ell$ are absent. Duality in this case
acts in a self-similar way exchanging the rank of the gauge groups 
$N_c \leftrightarrow N_f+n'k-N_c$ but not the form of the interactions. This 
generalizes the example presented in section \ref{subsec:quarticI}.

We can also understand this duality as an $m_i$-deformation of Seiberg duality in 
the ${\sf A}_{n'+1}$ case. From this point of view the magnetic theory has gauge
group $U(n'(N_f+k)-N_c)$ and superpotential
\beq
\label{hatoMspaf}
\widehat W_{n'+1}=\frac{g'_0}{n'+1}\tr {Y'}^{n'+1}
+\sum_{\ell=1}^{n'} M_\ell \tilde q {Y'}^{n'-\ell} q
+\sum_{i=1}^{N_f} m_i (M_2)^i_i
~.
\eeq
Repeating the analysis of appendix A in \cite{Elitzur:1997hc} we recover the 
dual description presented above. Many degrees of freedom are massive in 
the presence of the last term in \eqref{hatoMspaf} and by integrating them out 
we recover the dual gauge group $U(N_f+n'k-N_c)$ and the superpotential 
\eqref{hatoMspae}.

\subsection{Consequences for R-charges and RG flows}
\label{subsec:hatoMRcharges}

In the general $(n,n')$ case the condition for the existence of a supersymmetric
vacuum can be written in terms of the parameters $\lambda$, $x$ as
\beq
\label{hatoMRaa}
\lambda \leq \lambda^{\rm SUSY}_{n,n'}=\frac{nn'x}{x-n}
~.
\eeq
There is spontaneous supersymmetry breaking if $x>n$ and 
$\lambda_{n,n'}^{\rm SUSY}$ is the maximum value of the coupling.

Once again, this property shows that as we increase $\lambda$ in the undeformed 
theory, the R-charges $R_X$ and $R_{X'}$ decrease making more and more single 
trace operators relevant.  At sufficiently large coupling, beyond a critical value 
$\lambda^*_{n,n'}<\lambda_{n,n'}^{\rm SUSY}$, the operator 
$\alpha_n\tr X^{n+1}+\alpha'_{n'}\tr {X'}^{n'+1}$ becomes relevant in the 2-adjoint 
$\widehat {\sf O}_{\rm M}$ theory and drives it to a new set of IR fixed points. From 
this submanifold of fixed points further deformations with lower power (more relevant) 
single trace operators is possible. This picture implies a vast set of fixed submanifolds 
and RG flows connecting them. It would be interesting to obtain a better understanding 
of these theories.

In conclusion, we find that the $\NN=2$ Landau-Ginzburg-like CSM theories 
in this section exhibit a rich structure that bears many similarities with the structure
familiar from analogous four-dimensional $\NN=1$ SQCD theories. Several 
features, however, are new in three dimensions compared to the four-dimensional
case. For example, there are situations where the R-charges decrease more in three 
dimensions and the destabilization of the supersymmetric vacuum becomes more 
efficient. For instance, comparing the models with superpotential \eqref{hatoMspaa} 
in three and four dimensions we detect a region of parameters without a supersymmetric 
vacuum in three dimensions, but no such region in four dimensions \cite{Elitzur:1997hc}. 
Also, in our three-dimensional 2-adjoint $\widehat{\sf O}_{\rm M}$ theories we detect a 
large set of relevant deformations. A quick calculation of R-charges with $a$-maximization 
techniques in the four-dimensional $\widehat{\sf O}_{\rm M}$ theory reveals that both 
R-charges $R_X$ and $R_{X'}$ asymptote to a finite value around $\frac{1}{2}$ at large 
$x$ allowing for only a limited set of relevant deformations.

\section{Discussion}
\label{sec:discussion}

In this paper we considered $\NN=2$ Chern-Simons theories with $U(N_c)$ 
gauge group coupled to $N_f$ pairs of chiral superfields in the (anti)fundamental 
representation and zero, one or two chiral superfields in the adjoint. In the absence 
of superpotential interactions these theories are classically and quantum mechanically 
superconformal (at least within a range of parameters). Superpotential interactions 
can be added to generate RG flows towards new IR fixed points. The resulting theories 
can be viewed as three-dimensional generalizations of $\NN=(2,2)$ Landau-Ginzburg 
models in two dimensions and bear many similarities with $\NN=1$ (adjoint) SQCD 
theories in four dimensions. 

Using properties like spontaneous breaking of supersymmetry and Seiberg
duality we obtained:
\vspace{.2cm}

\noindent\vspace{.1cm}
(1) a list of semi-quantitative non-perturbative features of $U(1)_R$ symmetries,

\noindent\vspace{.1cm}
(2) a web of RG flows,

\noindent\vspace{.1cm}
(3) the postulation of new interacting fixed points partially admitting an
ADE classification,

\noindent\vspace{.1cm}
(4) interesting parallels between three- and four-dimensional gauge theories.

\vspace{.2cm}
\noindent
In the process, we argued for a set of new examples of Seiberg duality in 
three-dimensional CSM theories.

Our discussion provides an ample demonstration of the rich dynamics of 
Chern-Simons theories coupled to matter. It is of intrinsic interest to develop 
exact analytic methods that will allow us to study these properties further and 
beyond perturbation theory. 

One can think of several applications in string/M theory. For example, the 
low-energy dynamics of $N$ M2-branes in flat space is described by a quiver 
$U(N)\times U(N)$ CSM theory at level 1 with enhanced $\NN=8$ supersymmetry 
\cite{Aharony:2008ug}. This theory is strongly coupled. In this and other cases, 
CSM theories have a dual gravitational description in string or M theory. In all 
these cases, knowledge about the strong coupling dynamics of the CSM theories 
is useful not only per se but also for the dual four-dimensional quantum gravity 
description.

Some related more concrete questions that arise from this work are as follows.

\subsection*{What determines the R-symmetry in $\NN=2$ supersymmetric CSM theories?}

In three-dimensional superconformal field theories with $\NN=2$ supersymmetry 
we would like to know if there is an exact analytic method that determines the $U(1)_R$ 
symmetry. We have seen that this symmetry can receive large non-perturbative 
contributions. In four-dimensional gauge theories with the same amount of supersymmetry,
$i.e.$ $\NN=1$ supersymmetry, the exact $U(1)_R$ symmetry is determined
by $a$-maximization \cite{Intriligator:2003jj}. This method boils down to maximizing
a function $a$ that can be expressed as a linear combination of 't Hooft anomalies.
The method relies heavily on the ability to identify the candidate R-symmetries in the 
weakly coupled UV regime, so sometimes more than just weak coupling
data is needed to make this method practical. In some cases, this extra
information is provided by Seiberg duality \cite{Kutasov:2003iy}.

It is natural to ask if there is a similar principle at work in three-dimensional
gauge theories with $\NN=2$ supersymmetry, and more specifically
in $\NN=2$ CSM theories. Anomalies of continuous symmetries are absent
in three dimensions, so if there is an analog of $a$ in three dimensions it will 
be expressible in a different way. An interesting alternative to $a$-maximization
that works equally well in any dimension is $\tau_{RR}$-minimization 
\cite{Barnes:2005bm}. $\tau_{RR}$ is the coefficient of the two-point function of 
the $U(1)_R$ current. The exact superconformal $U(1)_R$ minimizes this coefficient. 
Unfortunately, there is no known efficient way of computing this coefficient analytically 
in interacting field theories. 

In four dimensions the function $a$ has been conjectured to be a good candidate
for a $c$-function \cite{Intriligator:2003jj,Kutasov:2003ux,Barnes:2004jj} (see,
however, \cite{Shapere:2008un}), $i.e.$ a function that is positive and monotonically 
decreasing along RG flows -- Zamolodchikov's $c$-function in two-dimensional 
quantum field theory being the prototype example \cite{Zamolodchikov:1986gt}. 
One wonders whether a tentative $a$ function in three dimensions would also be 
a good candidate for a $c$-function. Defining a $c$-function in three dimensions 
is a notoriously difficult problem (for work related to this problem see 
\cite{Anselmi:2002as,Anselmi:2002fk}).

Another interesting question is whether we can relate the $\NN=2$ CSM theories
to two-dimensional quantum field theories and thus obtain some answers to
the above questions from a two-dimensional perspective. The $\NN=2$ 
Chern-Simons theory with gauge group $G$ becomes, after integrating out the 
fermions, a bosonic Chern-Simons theory \eqref{introaa} at the shifted level 
\beq
\label{discussionaa}
k'=k-\frac{h}{2}\sgn(k)
\eeq
where $k$ is the $\NN=2$ CS level and $h$ the dual Coxeter number of the
gauge group $G$ \cite{Kao:1995gf}. This theory, which is a topological quantum 
field theory, is known to be equivalent to the (chiral) WZW model with gauge 
group $G$ and level $k'$ \cite{Witten:1988hf}.\footnote{Incidentally, from this 
perspective a three-dimensional interpretation of the two-dimensional central 
charge $c$ was given in \cite{Witten:1989ni}.}
We may ask whether a more general 2d/3d connection persists for CS 
theories coupled to matter, $e.g.$ when the three-dimensional CS theory 
is placed on a manifold with a two-dimensional boundary. This question
is also relevant for the dynamics of M2-branes ending on an M5-brane.
In fact, this may be a way to capture the dynamics of the self-dual string on 
the M5-brane worldvolume from the boundary dynamics of the CSM theory 
that lives on the M2-brane worldvolume.

\subsection*{RG flows and the ADE classification}

In section \ref{sec:hato} we presented a subclass of RG flows which appear
to admit an ADE classification. It would be interesting to establish the precise
range of parameters where these RG flows take place and prove the assertion
that the theories that describe the IR dynamics of these flows are superconformal.
In some cases, $e.g.$ the case of the ${\sf E}_n$ theories, the precise range of
the parameter $n$ needs to be determined. 

Assuming that a complete ADE classification takes place in the above subclass
of RG flows, it would be interesting to explore if there is a deeper connection
with other cases where the ADE classification occurs. For example, in 
two-dimensional $\NN=(2,2)$ theories the ADE superpotentials are special 
because they lead to the $\hat c<1$ minimal models in which all elements 
of the chiral ring are relevant operators \cite{Martinec:1989in,Vafa:1988uu}. 
Perhaps some of the well known results in two-dimensional $\NN=(2,2)$ 
superconformal field theories, $e.g.$ the superconformal Poincare polynomial 
of R-charges and other properties of critical points \cite{Cecotti:1992rm}, can 
be extended to the three-dimensional $\NN=2$ CSM theories presented in this 
paper.

\section*{Acknowledgements}
\noindent

I would like to thank Elias Kiritsis, David Kutasov, Georgios Michalogiorgakis and 
Carlos N\'u\~nez for interesting and enlightening discussions. This work has been 
supported by the European Union through an Individual Marie Curie Intra-European 
Fellowship. Additional support was provided by the ANR grant,  ANR-05-BLAN-0079-02, 
the RTN contracts MRTN-CT-2004-005104 and MRTN-CT-2004-503369, and the 
CNRS PICS {\#} 3059, 3747 and 4172.

\section*{Appendices}

\begin{appendix}

\vspace{-.3cm}
\section{Supersymmetries of a Hanany-Witten setup}
\label{app:HWsusies}

In this appendix we review the amount of supersymmetry preserved by the
configuration \eqref{SQCDquca}. We will find it convenient to work in M-theory
following the analysis of \cite{Kitao:1998mf}. Without the $N_f$ D5-branes the
M-theory version of \eqref{SQCDquca} is
\beq
\label{quaag}
\begin{array}{r c l}
N_c ~ M2~~&:&~~ 016
\\[1mm]
M5~~&:&~~ 012345
\\[1mm]
(1,k)~ M5'~\, &:&~~ 012\left[ {3 \atop 7} \right]_\theta  89
\end{array}
\eeq
The supersymmetry preserving conditions for these branes are
\begin{subequations}
\bea
\label{quaai}
M2~~&:& ~~~\Gamma_{016}\epsilon=\epsilon
~,\\
\label{quaaib}
M5~~&:&~~~\Gamma_{012345}\epsilon=\epsilon
~,\\
M5'~\, &:&~~~R\Gamma_{012345}R^{-1}\epsilon=\epsilon
\eea
\end{subequations}
where $R$ is the rotation matrix ($||$ denotes the 11th direction) 
\beq
\label{quaaj}
R=e^{\frac{\theta}{2}(\Gamma_{2||}+\Gamma_{37})-\frac{\pi}{4}(\Gamma_{48}+\Gamma_{59})}
\eeq
and $\Gamma_*$ are eleven-dimensional $\Gamma$-matrices. Ref.\ \cite{Kitao:1998mf} 
shows that this configuration preserves $\NN=2$ supersymmetry in three dimensions.

Now we add $N_f$ M5-branes along
\beq
\label{quaak}
\widetilde{M5}~~:~~~ 017 \left[ {4\atop 8} \right]_{\psi} \left[{5 \atop 9}\right]_{\psi} ||
~.
\eeq
These branes do not reduce the supersymmetry any further. Indeed, the supersymmetry
preserving condition for each of these branes is
\beq
\label{quaal}
\hat R \Gamma_{01789||} \hat R^{-1}\epsilon=\epsilon
\eeq
with the obvious $\psi$-dependent rotation matrix. Since
\beq
\label{quaam}
\Gamma_{0123456789||}=1 ~\Rightarrow ~ \Gamma_{01789||}=\Gamma_{016}
\Gamma_{012345}
\eeq
and
\beq
\label{quaan}
\Gamma_{012345}\hat R^{-1}=\hat R \Gamma_{012345}~, ~~
\Gamma_{016}\hat R=\hat R^{-1} \Gamma_{016}
\eeq
we deduce that eq.\ \eqref{quaal} follows from the pre-existing conditions
\eqref{quaai}, \eqref{quaaib}. Hence, no more supersymmetries are broken by the 
rotated D5-branes in the configuration \eqref{SQCDquca}.

\section{Evidence for Seiberg duality in the ${\sf D}_{n+2}$ theories}
\label{app:EviBrodie}

In this appendix we provide some evidence for the duality proposed in subsection 
\ref{subsec:hatoDduality}. The electric theory is a $U(N_c)$ $\NN=2$ CSM theory
at level $k$ coupled to $N_f$ pairs of (anti)fundamental chiral superfields $Q^i,\widetilde Q_i$
and two adjoint chiral superfields $X,X'$ with superpotential
\beq
\label{EviBrodieaa}
W_{{\sf D}_{n+2}}=\frac{g}{n+1}\tr X^{n+1}+g' \tr X{X'}^2
~.
\eeq
The proposed magnetic theory is a $U(3n(N_f+k)-N_c)$ $\NN=2$ CSM theory at level $k$ 
coupled to $N_f$ pairs of (anti)fundamental chiral superfields $q_i, \tilde q^i$, two adjoint 
chiral superfields $Y,Y'$ and $3nN_f^2$ gauge singlet superfields $(M_{\ell s})^i_j$
($\ell=1,\cdots,n,~s=1,2,3,~ i,j=1,\cdots, N_f$) with superpotential
\beq
\label{EviBrodieab}
\widetilde W_{{\sf D}_{n+2}}=\frac{\tilde g}{n+1}\tr Y^{n+1}+{\tilde g}' \tr Y{Y'}^2+
\sum_{\ell=1}^n\sum_{s=1}^3 M_{\ell s}\tilde q Y^{n-\ell}{Y'}^{3-s}q
~.
\eeq
The couplings in front of the meson superpotential interactions are kept implicit.

The mesonic part of the magnetic superpotential respects the global $SU(N_f)\times SU(N_f)$
symmetry and is such that the composite magnetic mesons
\beq
\label{EviBrodieac}
\widetilde \MM_{\ell s}\equiv \tilde q Y^{\ell-1}{Y'}^{s-1} q~, ~ ~ \ell=1,\cdots,n~, ~ ~
s=1,2,3
\eeq
are related to the elementary fields $M_{\ell s}$ by Legendre transform. The Legendre 
conjugate pairs are $M_{\ell s}$ and $\widetilde \MM_{n+1-\ell,4-s}$. When the 
corresponding term in $\widetilde W_{{\sf D}_{n+2}}$ is relevant we should include 
$M_{\ell s}$ in the spectrum of independent operators and drop $\widetilde \MM_{n+1-\ell,4-s}$.
Then, by Seiberg duality the elementary fields $M_{\ell s}$ are mapped to the electric
composite meson superfields $\widetilde Q Y^{\ell-1}{Y'}^{3-s} Q$.

Giving a complex mass to one of the quarks in the electric theory,
\beq
\label{EviBrodiead}
W_{{\sf D}_{n+2}} \to \frac{g}{n+1}\tr X^{n+1}+g' \tr X{X'}^2+m \widetilde Q_{N_f} Q^{N_f}
\eeq
we can integrate out the massive quarks $\widetilde Q_{N_f}$, $Q^{N_f}$ and flow
to a ${\sf D}_{n+2}$ theory with $N_f-1$ quark pairs.

On the magnetic side this deformation corresponds to the superpotential  
\beq
\label{EviBrodieae}
\widetilde W_{n+2} \to \frac{\tilde g}{n+1}\tr Y^{n+1}+{\tilde g}' \tr Y{Y'}^2+
\sum_{\ell=1}^n\sum_{s=1}^3 M_{\ell s}\tilde q Y^{n-\ell}{Y'}^{3-s}q
+m (M_{1,1})^{N_f}_{N_f}
~.
\eeq
The F-term equation for the meson $(M_{1,1})^{N_f}_{N_f}$ reveals that the 
Legendre conjugate composite meson $\widetilde \MM_{n,3}$ acquires a vacuum 
expectation value
\beq
\label{EviBrodiaf}
\tilde q^{N_f} Y^{n-1}{Y'}^2 q_{N_f}=-m
~.
\eeq
Solving the full set of F-term equations and the D-flatness conditions we obtain
non-vanishing expectation values for the quarks $\tilde q^{N_f}$, $q_{N_f}$
and the adjoint scalars $Y, Y'$. Before giving the solution we make a short
parenthesis to discuss explicitly the D-flatness conditions.

The relevant terms from the $\NN=2$ CSM Lagrangian are
\bea
\label{EviBrodieag}
\LL_D&=&\frac{k}{2\pi} D^\alpha_\beta \sigma^\beta_\alpha
-\sum_{i=1}^{N_f}\left( q^\dagger_{i\beta} (\sigma^2)^\beta_\alpha q^\alpha_i
+\tilde q^{\dagger i\beta}(\sigma^2)^\alpha_\beta \tilde q^i_\alpha
-q^\dagger_{i\beta}D^\beta_\alpha q^\alpha_i
+\tilde q^{\dagger i\alpha}D^\beta_\alpha \tilde q^i_\beta\right)
\nonumber\\
&&-[\sigma,Y]^{\dagger\beta}_\alpha [\sigma,Y]^\alpha_\beta
-[\sigma,Y']^{\dagger\beta}_\alpha [\sigma,Y']^\alpha_\beta
+Y^{\dagger \beta}_\alpha [D,Y]^\alpha_\beta
+{Y'}^{\dagger \beta}_\alpha [D,Y']^\alpha_\beta
~.
\eea
$\alpha,\beta$ are gauge indices for the fundamental representation and $\sigma, D$ 
are scalars in the $\NN=2$ vector mulitplet. The $D^\alpha_\beta$ act as Lagrangre 
multipliers whose equations of motion give
\bea
\label{EviBrodieai}
\sigma^\beta_\alpha=-\frac{2\pi}{k}
\left[ \sum_{i=1}^{N_f} \left( q^\dagger_{i\alpha} q^\beta_i
-\tilde q^{\dagger i\beta}\tilde q^i_\alpha\right)
+[Y^\dagger, Y]^\beta_\alpha
+[{Y'}^\dagger, Y']^\beta_\alpha
\right]
~.
\eea
Inserting this expression back into \eqref{EviBrodieag} we obtain the D-term
potential, which we require to vanish.

As an example, we consider the case with $n=2$, $N_c=10$, $N_f=2$ and 
$k=1$. The dual gauge group is $U(8)$. A solution that satisfies all the F-term 
equations and the D-flatness conditions has (for simplicity we set $\tilde g=\tilde g'=1$)
\begin{subequations}
\beq
\label{EviBrodieaj}
\tilde q^{N_f}_\alpha=-\left( \frac{m}{2}\right)^{1/5} \delta_{\alpha,1}~, ~~
q^\alpha_{N_f}=-\left( \frac{m}{2}\right)^{1/5} \delta^{\alpha,6}
~,
\eeq
\beq
\label{EviBrodieak}
Y=-\left(\frac{m}{8}\right)^{1/5}
\left(\begin{array}{cccccccc}
 0~~ & \sqrt{2}~~ & 0~~ & 0~~& 0~~& 0~~ & 0~~ & 0 \\
 0~~ & 0~~           & 0~~ & 0~~ & 1~~&0~~ & 0~~ & 0 \\ 
 0~~ & 0~ ~          & 0~~ & -1~~& 0~~&0~~ & 0~~ & 0 \\
 0~~ & 0~ ~          & 0~~ & 0~~ & 0~~ &0~~ & 0~~ & 0 \\
 0~~ & 0~ ~          & 0~~ & 0~~ & 0~~ & -\sqrt{2}~~ &0~~ & 0 \\
 0~~ & 0~ ~          & 0~~ & 0~~ & 0~~ &0~~ & 0~~ & 0 \\
 0~~ & 0~ ~          & 0~~ & 0~~ & 0~~ &0~~ & 0~~ & 0 \\
 0~~ & 0~ ~          & 0~~ & 0~~ & 0~~ &0~~ & 0~~ & 0 
\end{array}\right)
\, , ~ 
\eeq
\beq
\label{EviBrodieal}
Y'=-\left(\frac{m}{8}\right)^{1/5}
\left(\begin{array}{cccccccc}
 0~~ & 0~ & \sqrt{2}~~ & 0~~& 0~~& 0~~ & 0~~ & 0 \\
 0~~ & 0~ ~          & 0~~ & 1~~ & 0~~&0~~ & 0~~ & 0 \\ 
 0~~ & 0~ ~          & 0~~ & 0~~& -1~~&0~~ & 0~~ & 0 \\
 0~~ & 0~ ~          & 0~~ & 0~~ & 0~~ &\sqrt{2}~~ & 0~~ & 0 \\
 0~~ & 0~ ~          & 0~~ & 0~~ & 0~~ & 0~~ &0~~ & 0 \\
 0~~ & 0~ ~          & 0~~ & 0~~ & 0~~ &0~~ & 0~~ & 0 \\
 0~~ & 0~ ~          & 0~~ & 0~~ & 0~~ &0~~ & 0~~ & 0 \\
 0~~ & 0~~           & 0~~ & 0~~ & 0~~ &0~~ & 0~~ & 0 
\end{array}\right)
~.
\eeq
\end{subequations}
The solution has the same form as in the four-dimensional ${\sf D}_{n+2}$ magnetic
theory \cite{Brodie:1996vx}.

In the general case the solution takes the form
\bea
\label{EviBrodieam}
&&\tilde q^{N_f}_\alpha \sim \delta_{\alpha,1}~, ~~
q^\alpha_{N_f} \sim \delta^{\alpha,3k}
~, ~~
\nonumber\\
&&Y^\alpha_{\alpha+1}\neq 0~, ~~ {\rm but}~~ Y^k_{k+1}=Y^{2k}_{2k+1}=0~, ~~
{\rm and}~~ Y^k_{2k+1}\neq 0
~,\\
&&{Y'}^\alpha_{\alpha+k} \neq 0
\nonumber
\eea
with all other elements zero. The precise values of the non-vanishing elements
can be determined as above by solving the F and D-flatness equations.

The above vacuum expectation values Higgs the gauge group from 
\beq
\label{EviBrodiean}
U(3n(N_f+k)-N_c) \to U(3n(N_f+k-1)-N_c)
~.
\eeq
At the same time the $q_{N_f}$ and $\tilde q^{N_f}$ quarks are eaten by the gauge
group and disappear. The adjoint fields $Y$, $Y'$ break into smaller $U(3n(N_f+k-1)-N_c)$
matrices and $6n$ fundamentals. $3n-1$ of these fundamentals are eaten by the
Higgs mechanism and $3n+1$ of them become massive. In the IR we recover 
the theory which is expected to be the magnetic dual to the mass deformed electric
theory \eqref{EviBrodiead}.

\end{appendix}


\end{document}